\documentclass[10pt,twocolumn,floatfix,superscriptaddress,amsmath,showkeys,aps,prb,eqsecnum]{revtex4-2}
\usepackage{bm}
\usepackage{xcolor}
\usepackage{hyperref}
\usepackage{makeidx}
\usepackage{amsmath}
\usepackage{graphicx}
\usepackage{dcolumn}
\usepackage{float}
\usepackage{amsfonts}
\usepackage{amssymb}
\usepackage{color}
\usepackage{array}

\newcommand {\vk} {\bf k}
\newcommand {\vQ} {\bf Q}
\newcommand {\vR} {\bf R}
\newcommand \sa {\sigma}

\newcommand \up {\uparrow}
\newcommand \pu {\downarrow}
\newcommand \epk[1] {\epsilon_{{\bf k}}^{#1}}

\newcommand \epks[2] {\epsilon_{{\bf k}{#2}}^{#1}}
\newcommand \epkqs[2] {\epsilon_{{\bf k+Q}{#2}}^{#1}}
\newcommand \epkms[2] {\epsilon_{{\bf -k}{#2}}^{#1}}
\newcommand \epkqms[2] {\epsilon_{{\bf -k-Q}{#2}}^{#1}}

\newcommand \ckm[1]  {c_{#1}^{\dag}}
\newcommand \ck[1]     {c_{#1}}
\newcommand \fkm[1] {d_{#1}^{\dag}}
\newcommand \fk[1] {d_{#1}}
\newcommand \med[1] {\langle{#1}\rangle}

\hypersetup{
pdfauthor = {Lopes, Reyes, Continentino and Thomas},
pdftitle = {Interplay between charge density wave and superconductivity in multi-band systems with interband Coulomb interaction},
}

\begin{document}

\title{Interplay between charge density wave and superconductivity in multi-band systems with inter-band Coulomb interaction}

\author{Nei Lopes}
\email{nlsjunior12@gmail.com}
\affiliation{Departamento de F\'{\i}sica Te\'orica, Universidade do Estado do Rio de Janeiro, Rua S\~ao Francisco Xavier 524, Maracan\~a, 20550-013, Rio de Janeiro, RJ, Brazil}
\author{Daniel Reyes}
\email{daniel@cbpf.br}
\affiliation{Instituto Militar de Engenharia - Pra\c{c}a General Tib\'{u}rcio, 80, 22290-270, Praia Vermelha, Rio de Janeiro, Brazil}
\author{Mucio A. Continentino}
\affiliation{Centro Brasileiro de Pesquisas F\'{\i}sicas, Rua Dr. Xavier Sigaud 150, Urca, 22290-180, Rio de Janeiro , Brazil}
\author{Christopher Thomas}
\email{Present address: Instituto de F\'{\i}sica, UFRGS, 91501-970 Porto Alegre, Brazil}
\affiliation{Departamento de F\'isica, Universidade Federal Rural do Rio de Janeiro, 23897-000, Serop\'edica, Rio de Janeiro, Brazil}

\date{\today}

\begin{abstract}
In this work we study the competition or coexistence  between charge density wave (CDW) and superconductivity (SC) in a two-band model system in a square lattice. One of the bands has a net attractive interaction ($J_d$) that is responsible for SC. The model includes on-site Coulomb repulsion between quasi-particles in different bands ($U_{dc}$) and the hybridization ($V$) between them. We are interested in describing inter-metallic systems with a $d$-band of moderately correlated electrons, for which a mean-field approximation is adequate, coexisting with a large $sp$-band. For simplicity, all interactions and the hybridization $V$ are considered site-independent. We obtain the eigenvalues of the Hamiltonian numerically and minimize the free energy density with respect to the relevant parameters to obtain the phase diagrams as function of $J_d$, $U_{dc}$, $V$, composition ($n_{\mathrm{tot}}$) and the relative depth of the bands ($\epsilon_{d0}$).  We consider two types  of superconducting ground states coexisting with the CDW. One is a homogeneous ground state and the other is a pair density wave where the SC order parameter has the same spatial modulation of the CDW. Our results show that the CDW and SC orders compete, but depending on the parameters of the model these phases may coexist. The model reproduces most of the experimental features of high dimensionality ($d>1$) metals with competing CDW and SC states, including the existence of first and second-order phase transitions in their phase diagrams.
\end{abstract}


\maketitle

%%%%%%%%%%%%%%%%%%%%%%%%%%%%%
\section{Introduction}

Quantum phase transitions to or from a superconducting state driven by pressure, doping or magnetic field have long been one of the most intriguing and extensively studied phenomena in solid-state physics. These transitions can occur directly from a normal Fermi liquid state, from a non-Fermi liquid or from another phase with broken symmetry. An interesting example of the latter is the Charge Density Wave (CDW)-Superconductor (SC) phase transition~\cite{Morosan2006,Kusmartseva2009,Gruner2017}. Both SC and CDW are symmetry breaking phases characterized by an energy gap in the single-particle spectrum and with order parameters representing the condensation of electron–electron or electron–hole pairs, respectively.

The CDW phase corresponds to a condensate with periodic modulation of the electron density, often found in low-dimensional transition metal dichalcogenides (TMD's)~\cite{Morosan2006,Zhao2007,Wilson1975,Kusmartseva2009,Sipos2008,Yang2012,Pyon2012,Fang2013,Kamitani2016,Kudo2016,Heil2017,SaintPaul2020}, A15 compounds~\cite{Chu1974,Chu1974a,Testardi1975,Tanaka2010}, cuprates~\cite{BussmannHolder1992,Birgeneau1987,Wakimoto2004}, Ni- and Fe-based SCs~\cite{Cruz2008,Yoshizawa2012,Niedziela2011,Kudo2012,Hirai2012}, perovskites~\cite{Kang2011}, quasi-skutterudite SC~\cite{Kase2011,Wang2012,Klintberg2012,Zhou2012,Biswas2014,Goh2015}, intercalated graphite CaC$_6$~\cite{Gauzzi2007}, organic compounds~\cite{Lubczynski1996,Wosnitza2001}, and sulfuride-based compounds at very high pressure~\cite{Degtyareva2007}. It is surprising to find in this list also three-dimensional materials, since there were no expectations to find CDW in high dimensions~\cite{Klintberg2012}.

In many cases these systems exhibit SC in normal conditions~\cite{Yu2015,Cheung2018,Goh2015} or upon tuning
with non-thermal parameters, such as physical pressure~\cite{Wilson1975,Kusmartseva2009,Testardi1975,Coleman1973,Edwards1994,Gabovich2001,Gabovich2002,Klintberg2012} and chemical pressure from partial atomic substitution~\cite{Klintberg2012,Yu2015,Goh2015}. Typically,
the SC is found to be abruptly enhanced upon suppressing the CDW order with a phase diagram that remarkably resembles those
of the cuprates~\cite{Dagotto2005,Stewart2017}, heavy fermions~\cite{Mathur1998,Gegenwart2008} and  iron-based superconductors~\cite{Paglione2010,Ishida2009,Hashimoto2012,Shibauchi2014}, where the competing order of SC is the charge order, instead of the antiferromagnetic spin order. 

In principle, these phenomena can be approached within the framework of Bardeen-Cooper-Schrieffer (BCS) theory~\cite{Bardeen1957,Bardeen1957a} due to the nodeless nature of the superconducting gap function~\cite{Kase2011,Hayamizu2011,Wang2012,Zhou2012,Biswas2014}, as indicated by the specific heat temperature dependence, and the ratios $2\Delta/k_B T_{\mathrm{SC}}$ and $\Delta C/\gamma T_{\mathrm{SC}}$ close to the expected value in BCS theory~\cite{Bardeen1957,Bardeen1957a}.

There is a longstanding question concerning the competition, coexistence, or even cooperation between CDW and SC, as well as about the nature of the CDW phase transition. For the latter, several works have reported a first-order structural phase transition where pressure initially suppresses  the CDW, but enhances the superconducting transition temperature $T_{\mathrm{SC}}$~\cite{Shen2020}, whereas others find a second-order phase transition that extrapolates to  a structural quantum critical point (SQCP), around which a dome-shaped variation of $T_{\mathrm{SC}}$ is found~\cite{Yu2015,Cheung2018,Poudel2016}. 
For instance, for the compound LaPt$_2$Si$_2$~\cite{Shen2020} with transition temperature $T_{\mathrm{SC}}$=1.87 K, and a superlattice structural transition at $T^{\star}$=76  K, it is suggested that the occurrence of a SC dome-like can be accounted for, within the BCS theory, assuming there is a maximum in the density of states $N(E_F )$ upon the closure of the CDW gap. Then, the evolution of $T_{\mathrm{SC}}$ under pressure  is likely driven by the variation of $N(E_F )$, with a sudden disappearance of CDW order, which indicates that there is a first-order structural transition suggesting the lack of a QCP in this material~\cite{Shen2020}. On the other hand, the cubic superconducting inter-metallic systems Lu(Pt$_{1-x}$Pd$_x$)$_2$In and (Sr$_{1-x}$Ca$_x$)$_3$Ir$_4$Sn$_{13}$ present second-order CDW phase transitions, under chemical doping and pressure, respectively~\cite{Carneiro2020}.

In this work we investigate the interplay between CDW and SC orders in a square lattice two-band model. The bands have different effective masses and the CDW phase arises from inter-band Coulomb correlations. On the other hand, the SC is due to a local intra-band attractive interaction. The interactions and the hybridization between bands are considered, for simplicity, to be $\vk$-independent. We also study and compare the case where the SC can be modulated. Our results show that this case does not support coexistence.

We are particularly interested in inter-metallic compounds and their alloys~\cite{SaintPaul2020}, like the layered SCs SrPt$_2$As$_2$ and LaPt$_2$Si$_2$~\cite{Shen2020,Kim2015}, the cubic superconducting systems Lu(Pt$_{1-x}$Pd$_x$)$_2$In and (Sr$_{1-x}$Ca$_x$)$_3$Ir$_4$Sn$_{13}$~\cite{Carneiro2020}, the cubic Heusler alloys Lu(Pt$_{1-x}$Pd$_x$)$_2$In~\cite{Gruner2017}, and  the large family of superconducting stannides  with composition A$_3$T$_4$Sn$_{13}$, where A = La; Sr; Ca and T = Rh; Ir~\cite{Biswas2015}. These systems have in common a rather narrow $d$-band with moderate electronic correlations (as compared with the $f$-bands in heavy fermions) coexisting with large $sp$-bands~\cite{Ban2017}. Their multi-band character, the existence of moderate electronic correlations, the BCS nature of their superconducting ground states and the interplay of first and second order transitions were guides to build our model and for the approach we adopted. 

The many-body problem posed by our Hamiltonian is solved within a Hartree-Fock (HF) mean-field approximation. We use Nambu's spinor representation to write the Hamiltonian in matrix form. We obtain its eigenvalues numerically and minimize the free energy density with respect to the relevant variables to obtain the phase diagrams as  functions of  parameters, such as the strength of the CDW and SC interactions, hybridization, total number of particles and relative depth of the bands.

We find that there is an intrinsic competition between CDW and SC orders, but depending on the parameters of the model these phases may coexist. However,  for modulated SC,  coexistence is totally suppressed. Our results show that for half-filling, the fine-tuned point in the phase diagram where superconductivity disappears  is that associated with the maximum value of the CDW critical temperature for very small inter-band Coulomb correlations. The CDW phase emerges around half-filling ($n_{\mathrm{tot}}=2.0$), as expected~\cite{Brydon2005}. It spreads out in the phase diagram as the Coulomb inter-band interaction increases. Away from half-filling we obtain coexistence of phases in qualitative agreement with experimental results for compounds with a discontinuous vanishing of the CDW order, although with a low temperature persistent SC phase. In addition, a reentrant behavior of $T_{\mathrm{CDW}}$ for large values of $U_{dc}$ is observed, which is a direct signature of a first-order phase transition. 

We also identify another feature that acts in detriment of  the coexistence of phases within this model, that is shifting the relative center of the bands ($\epsilon_{d0}$). This can be relevant when doping the $d$-elements with others of different rows in the periodic table or when applying pressure in the system. This suppression arises since SC is robust to this shift, while the CDW phase is very sensitive to it. Therefore, we show how $V$, $\epsilon_{d0}$, band-filling and strength of the correlations affect the competition between CDW and SC in multi-band compounds.

The paper is organized as follows: In Sec.~\ref{Model}, we present the two-band model with its main features  to study the competition or coexistence between CDW and SC in inter-metallic systems. For simplicity, we consider local hybridization and interactions.  We also briefly describe the Hartree-Fock approach that allows to solve the many-body problem. In Sec.~\ref{Results}, we show our results for finite-temperature phase diagrams as functions of the strength of the interactions, hybridization, band-filling and relative depth of the bands. We investigate both cases, of half-filling and away from half-filling occupations. In addition, we discuss results for the competition between CDW and a homogeneous SC and with a pair density wave order, the latter with the same spatial modulation of the CDW. Finally, in Sec.~\ref{conclusions}, we point out and summarize our main results.

%%%%%%%%%%%%%%%%%%%%%%%%%%%%%%%%%%%%
\section{Model}\label{Model}

In order to analyze the interplay between SC and CDW in inter-metallic compounds and their alloys, we consider a two-dimensional, two-band model in a square lattice with on-site inter-band Coulomb repulsion ($U_{dc}$) between electrons in different bands. Previously,  one-band models have been extensively used for studying the interplay between such phases~\cite{Balseiro1979,Scalettar1989,Vekic1992,Das2008}. 

The quasi-particles have different effective masses in distinct bands. The band with larger effective mass, which we refer generically as the $d$-band  has a local attractive interaction $J_d$ among its quasi-particles. This band hybridizes with a wide band of $c$-electrons by means of a real, symmetric, site-dependent hybridization. 

The real-space Hamiltonian of the model is given by,

\begin{align}\label{ref:hamiltonian}
  H = & \quad t_c\sum_{<ij>\sa}\ckm{i\sa}\ck{j\sa}+t_d\sum_{<ij>\sa}\fkm{i\sa}\fk{j\sa} \notag \\
  &+\sum_{i\sa}V_{ij}(\ckm{i\sa}\fk{j\sa}+\fkm{i\sa}\ck{j\sa})\notag\\
  &+U_{dc}\sum_{i}n^d_in^c_i+ J_d\sum_i \fkm{i\up}\fk{i\up}\fkm{i\pu}\fk{i\pu} %\notag \\
\end{align}
where $t_c$ ($t_d$) are the nearest-neighbor hopping energies of $c(d)$-electrons, $\ckm{i\sa}$ ($\ck{i\sa}$) and $\fkm{i\sa}$ ($\fk{i\sa}$) are creation (annihilation) operators associated with the $c$ and $d$-electrons with spin $\sigma$, respectively. The electrons in different bands are hybridized by means of a matrix with site-dependent elements $V_{i}$. $U_{dc}$ is the on-site inter-band repulsive interaction ($U_{dc}>0$) among the $d$ and $c$-electrons~\cite{Miyake2000}, and $n^{d(c)}_{i}$ are the occupation numbers. We define the number of particles for each band as $n^{d(c)}=n^{d(c)}_{\up}+n^{d(c)}_{\pu}$, where $n^{d(c)}_{\up} = n^{d(c)}_{\pu}$ since we are interested in paramagnetic solutions. The last term in Eq.~(\ref{ref:hamiltonian}) describes an on-site effective attraction between $d$-electrons  ($J_d < 0$) and is responsible for superconductivity. Notice that this term also describes antiferromagnetic (AFM), $xy$-type  exchange interactions between these electrons, such that  magnetic and superconducting ground states are in competition. In this work we are only interested in the latter. We have also neglected in this interaction an Ising term that when decoupled in the superconducting channel leads to $p$-wave pairing that is not considered here.

We consider the possibility of the CDW state by allowing for a periodic modulation of the average values of the occupation numbers as follows~\cite{Brydon2005}, 
\begin{align}
  \med{n_{i}^d}&=n^d+\delta^d\cos{(\vQ\cdot\vR_i)}\,,\\
  \med{n_{i}^c}&=n^c+\delta^c\cos{(\vQ\cdot\vR_i)}\,,
\end{align}
where $\med {n_{i}^d}$ ($\med{n_{i}^c}$) is the average number of ${d}(c)$-electrons per site, the modulation wave-vector $\vQ=$$(\pi/\mathrm{a},\pi/\mathrm{a})$ is the nesting vector~\cite{Chikina2020} and $\mathrm{a}$ is the lattice parameter. Moreover, the  order parameters of the CDW phase are $\delta^{d}$ and $\delta^{c}$ for $d$ and $c$-electrons, respectively. Here, we neglect an excitonic phase, that is, we consider $\med{\fkm{i\sa}\ck{i\sa}}=0$.

For a homogeneous system, in the absence of charge ordering, a spatial uniform solution is assumed and within the standard mean-field (MF) approach $\delta^d=\delta^c=0$ for all values of the Coulomb interaction~\cite{Brydon2005}.

We can decouple the interaction terms within a MF approximation and write the Hamiltonian, Eq.~(\ref{ref:hamiltonian}), in \textit{momentum} space as,

\begin{align}\label{ref: HF_hamiltonian}
  H_{MF}=&\sum_{\vk\sa}{'}\epk{c}\ckm{\vk\sa}\ck{\vk\sa}+\sum_{\vk\sa}{'}\epks{d}{\sa}\fkm{\vk\sa}\fk{\vk\sa}\notag \\
  &+\delta^c\sum_{\vk\sa}{'}\left(\fkm{\vk+\vQ\sa}\fk{\vk\sa}+\fkm{\vk\sa}\fk{\vk+\vQ\sa}\right) \notag \\
  &+\delta^d\sum_{\vk\sa}{'}\left(\ckm{\vk+\vQ\sa}\ck{\vk\sa}+\ckm{\vk\sa}\ck{\vk+\vQ\sa}\right) \notag \\
  &+\sum_{\vk\sa}{'}\left(V_{\vk}\ckm{\vk\sa}\fk{\vk\sa}+V_{\vk}^*\fkm{\vk\sa}\ck{\vk\sa}\right)\notag \\
  &+\sum_{\vk}{'}\left(\Delta_{\vk}^d\fkm{\vk\up}\fkm{-\vk\pu}+\Delta_{\vk}^{d*}\fk{-\vk\pu}\fk{\vk\up}\right)
  +\mathcal{C}_1\,
\end{align}
where $\Delta_{\vk}^d$ is the superconducting order parameter and,
\begin{align}
\epk{}&=-2t_c\left[\cos(k_x a)+\cos(k_y a)\right]\,,\\
\epk{c}&\equiv\epk{}+U_{dc}n^d-\mu\,,\label{eq:ekcufc}\\
\epk{d}&\equiv\gamma\epk{}+U_{dc}n^c-\mu+\epsilon_{d0}\,,\label{eq:ekfufc}\\
\delta^d&\equiv \frac{U_{dc}}{N}\sum_{\vk}{'}\left(\med{\fkm{\vk+\vQ\sa}\fk{\vk\sa}}+\med{\fkm{\vk\sa}\fk{\vk+\vQ\sa}}\right)\,,\\
\delta^c&\equiv \frac{U_{dc}}{N}\sum_{\vk}{'}\left(\med{\ckm{\vk+\vQ\sa}\ck{\vk\sa}}+\med{\ckm{\vk\sa}\ck{\vk+\vQ\sa}}\right)\,,\\
\Delta_{\vk}^d&\equiv J_{d}\med{\fk{-\vk\pu}\fk{\vk\up}}\,,\\
\mathcal{C}_1&=-N U_{dc}n^dn^c-N\delta^d\delta^c/U_{dc}-N|\Delta^d|^2/J_{d} \notag \\ &+N\mu(n^c+n^d),\label{eq:c1}
\end{align}
with $\gamma=t_d/t_c$ the inverse ratio of effective masses and $\mu$ the chemical potential. We consider a relative shift between the band centers given by $\epsilon_{d0}$  and  remark that $\sum_{\vk}^{'}$ represents a sum over the reduced Brillouin zone that is halved  due to the CDW instability. The bands are described by the dispersion relations $\epsilon_{\vk}^{c}$ and $\epsilon_{\vk}^{d}$ for $c$ and $d$-electrons in a convenient notation. Note that these dispersions are spin ($\sigma$) independent since we are not interested in magnetic solutions.

In the Hamiltonian, Eq.~(\ref{ref: HF_hamiltonian}), the chemical potential $\mu$ has to be adjusted self-consistently when we fix the total band-filling, i.e., $n_{\mathrm{tot}}=\sum_i \big(\med{n_{i}^{d}}+\med{n_{i}^c}\big)$ at different values.

Notice that Eq.~(\ref{ref: HF_hamiltonian}) can be written in the form of a matrix using the spinor Nambu basis,
\begin{equation}\label{ref: Nambu spinor representation}
\Psi^{\dag}\! =\! \left(
\!  \ckm{\vk\up}\!, \ckm{\vk+\vQ\up}\!, \fkm{\vk\up} \!, \fkm{\vk+\vQ\up}\!, \ck{-\vk\pu}\!, \ck{-\vk-\vQ\pu}\!,  \fk{-\vk\pu} \!, \fk{-\vk-\vQ\pu} \right) \\
\end{equation}
such that,
\begin{align}
  H_{MF}&=\sum_{\vk}{'}\Psi^{\dag}_{\vk}M\Psi_{\vk}+\mathcal{C}_1+\mathcal{C}_2\,
\end{align}
where $\mathcal{C}_2=\sum_{\vk}{'}\left(\epks{c}{}+\epkqs{c}{}+\epks{d}{}+\epkqs{d}{}\right)$ and
{\footnotesize
\begin{align}\label{ref:M}
  & M= \notag \\ 
  & \begin{pmatrix}
  \epks{c}{} & \delta^d & V_{\vk} & 0 & 0 & 0  & 0 & 0   \\
  \delta^d & \epkqs{c}{} & 0 & V_{\vk+\vQ} & 0 & 0 & 0 & 0  \\
  V_{\vk}^* & 0 & \epks{d}{} & \delta^c & 0 & 0 & \Delta_{\vk}^d & 0 \\
  0 & V_{\vk+\vQ}^* & \delta^c & \epkqs{d}{} & 0 & 0 & 0 & \Delta_{\vk+\vQ}^{d} \\
  0 & 0 & 0 & 0 & -\epkms{c}{} & -\delta^{d} & -V_{-\vk}^* & 0 \\
  0 & 0 & 0 & 0 & -\delta^{d} & -\epkqms{c}{} & 0 & -V_{-\vk-\vQ}^* \\
  0 & 0 & \Delta_{\vk}^{d*} & 0 & -V_{-\vk} & 0 & -\epkms{d}{} & -\delta^c \\
  0 & 0 & 0 & \Delta_{-\vk-\vQ}^{d*} & 0 & -V_{-\vk-\vQ} & -\delta^c& -\epkqms{d}{} 
   \end{pmatrix}
\end{align}
}

There are eight eigenvalues $E_{m \vk}$ but only four of them are independent since, $E_{5 \vk}=-E_{1 \vk}$, $E_{6 \vk}=-E_{2 \vk}$, $E_{7 \vk}=-E_{3 \vk}$ and $E_{8 \vk}=-E_{4 \vk}$. Therefore, the diagonalized Hamiltonian ($H_{\mathrm{diag}}$) can be written as, 
\begin{align}
  H_{\mathrm{diag}}=&\sum_{\vk}{'}\sum_{m=1,2,3,4}E_{m \vk}\left(\alpha^{\dag}_{m \vk}\alpha_{m \vk}+\beta^{\dag}_{m \vk}\beta_{m \vk}\right)\notag \\
 & +\mathcal{C}_1+\mathcal{C}_2+\mathcal{C}_3\,,
\end{align}
where $(\alpha,\beta)^{\dag}_{m \vk}$ and $(\alpha,\beta)_{m \vk}$ are new operators given by a linear combination of the original band operators $(c,d)^{\dag}$ and $(c,d)$, and 
\begin{align}
\mathcal{C}_3=-\sum_{\vk}{'}\left(E_{1\vk}+E_{2\vk}+E_{3\vk}+E_{4\vk}\right)\,.
\end{align}

The free energy density is calculated as follows~\cite{santos2010},
\begin{align}\label{free_energy}
  F=&-2T\sum_{\vk}{'}\sum_m\ln{[1+\exp{(-\beta E_{m\vk})}]} \notag \\
  &+\mathcal{C}_1+\mathcal{C}_2+\mathcal{C}_3\,
\end{align}
where $\beta=1/(k_B T)$ with $k_B$  the Boltzmann constant and $T$ the absolute temperature.

We remark that we consider a commensurate periodic modulation of the crystal lattice with $\vQ=(\pi/$a$,\pi/$a$)$ that doubles the Wigner-Seitz cell. Then, we divide by 2 the sum in $\vk$-space, where the original Brillouin zone sum is done using the special points technique developed by Chadi-Cohen~\cite{Chadi1973}.

%%%%%%%%%%%%%%%%%%%%%%%%%%%%%%%%%%
\section{Results}\label{Results}

In this section we present and discuss our results for the finite temperature phase diagrams as functions of  the  strengths of the CDW and SC interactions, the total number of particles per site, the hybridization and the relative depth of the bands. We point out that the last two parameters can be modified by applying external pressure in the system and, that for simplicity we consider an $\vk$-independent hybridization, such that $V_{\vk}=V_{\vk+\vQ}=V$. Initially, we consider  the nesting condition for the SC order parameter, i.e., $\Delta^{d}_{\vk}=\Delta^{d}_{\vk+\vQ}\equiv\Delta^{d}$ and later we investigate the effects of a modulation of this parameter in the coexistence between SC and the CDW states. 

The phase diagrams of the model can be obtained by numerical minimization of the free energy density, Eq.~(\ref{free_energy}), with respect to the parameters of the model, that is, solving numerically the self-consistent equations~\cite{Costa2018},
\begin{align}\label{ref: minimization}
  \frac{\partial F}{\partial \mu}=  \frac{\partial F}{\partial n^d}=  \frac{\partial F}{\partial \delta^d}=  \frac{\partial F}{\partial \delta^c}=\frac{\partial F}{\partial \Delta^d}\equiv0\,.
\end{align}

Energies will be renormalized by the hopping $t_c$, i.e., we take  $t_c=1.0$. Moreover, we take the lattice parameter $\mathrm{a}=1.0$, $k_B=1.0$ and  for the inverse ratio of effective masses $\gamma=0.4$. The latter is adequate for the $sp$ and $d$-bands of the inter-metallic compounds and their alloys that we are interested.

Since the values of CDW order parameters $|\delta^{c}|$ and  $|\delta^{d}|$ vanish at the same critical temperature and present similar variations  with $U_{dc}$ and other parameters, we present just the behavior of $|\delta^{c}|$ in the next figures, except in Fig.~\ref{fig:parametersxef0_t001} where both are compared explicitly. We start with results for the half-filling case, $n_\mathrm{tot}=2.0$,  and then  discuss different values of composition.

%%%%%%%%%%%%%%%%%%%%%%%%%%%%%%%%%%%%%%%%%%%%%%%%%%%%%%%%%%%%%%%%%%%%%%%%%%%
\subsection{Half-filling \texorpdfstring{$n_{\mathrm{tot}}=2.0$}{ntot=2}}

%%%%%%%%%%%%%%%%%%%%%%%%%%%%%%%%%%%%%%%%%%%%%%%%%%%%%%%%%%%%%%%%%%%%%%%%%%%%%
\subsubsection{Effects of \texorpdfstring{$J_{d}$ and $U_{dc}$}{Jd and Udc} in the temperature dependent phase diagrams}

\begin{figure}[b]\centering
  \includegraphics[width=0.92\columnwidth]{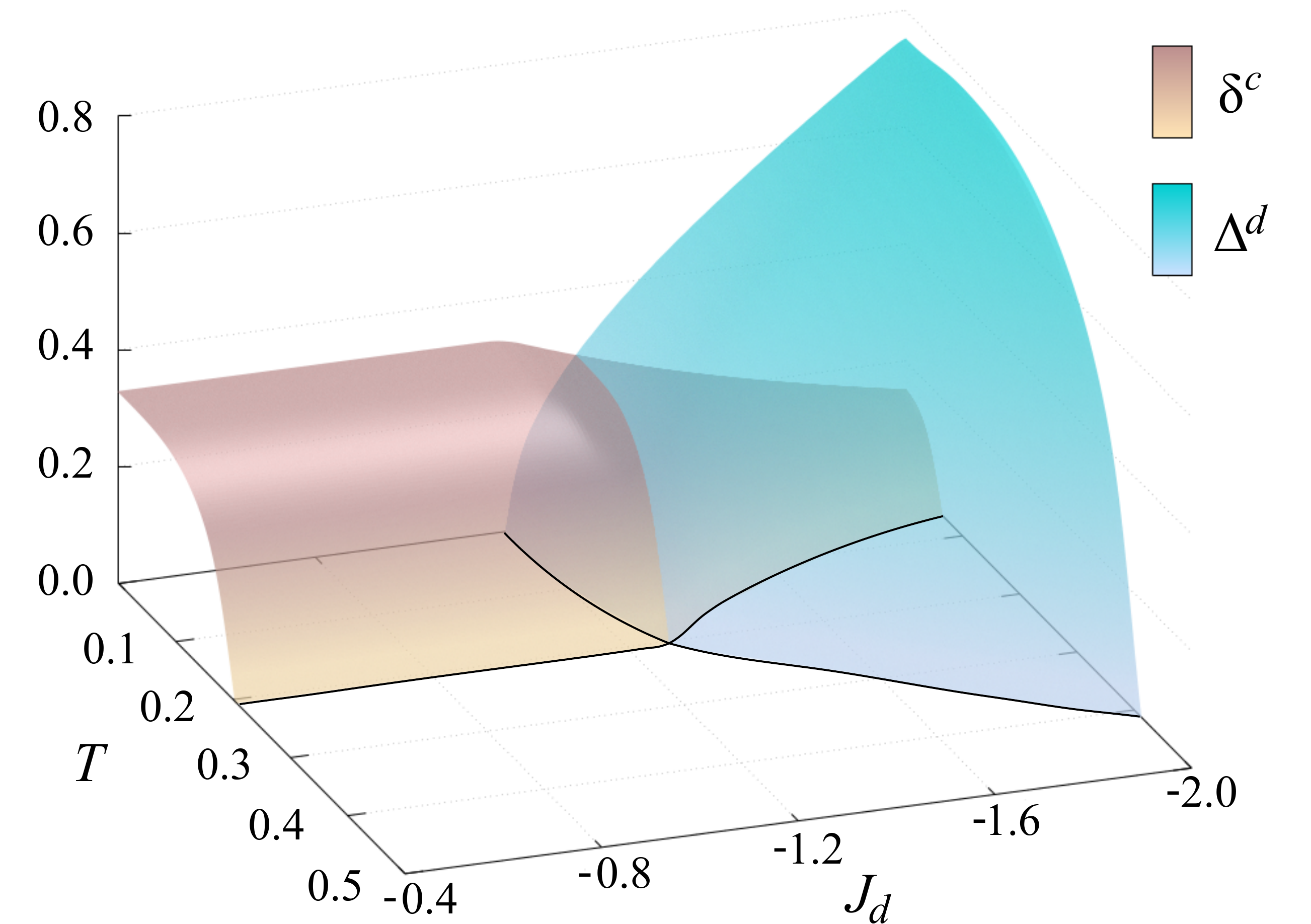}
  \caption{\label{fig:jxtvarufc} Plot of $\delta^{c}$ and $\Delta^d$ as functions of $J_{d}$ and $T$, for $U_{dc}=0.8$, $n_{\mathrm{tot}}=2.0$, $V=0.0$  and $\epsilon_{d0}=0.0$. The CDW order parameter $\delta^{c}$ does not vary with $J_d$ in the absence of the SC phase, being nearly constant for a fixed $T$. The coexistence region emerges for $|J_{d}| \approx 1.2 > U_{dc}$. There is an intrinsic competition between different  orders in the coexistence region and as we increase $\left|J_d\right|$ in the SC phase, $\delta^{c}$ is  suppressed asymptotically. All phase transitions in this figure are second-order (continuous solid line).} 
\end{figure}

In Fig.~\ref{fig:jxtvarufc}, we show the variation of the CDW and SC order parameters, $\delta^c$ and $\Delta^d$, respectively, as a function of temperature ($T$) and the attractive interaction ($J_d$) for $U_{dc}=0.8$. One can see the competitive nature of the CDW and SC orders in the coexistence region. In the region of the phase diagram where there is only CDW order, its order parameter does not vary with $J_d$, being nearly constant for a fixed $T$. However, when the superconducting state sets in, $\delta^{c}$ decreases in the coexistence region as we increase $\left|J_d\right|$. 

For small temperatures, in the presence of the CDW state, superconductivity requires a minimum value of the attractive interaction to appear. The scale for this critical value is set by the Coulomb inter-band repulsion so that, in general,  $\left|J_d\right|> U_{dc}$.   Notice that as we keep increasing $\left|J_{d}\right|$, the SC order becomes dominant at expenses of the CDW state, as $\delta^{c}$ vanishes asymptotically.

One important result to be considered is that there is no reciprocity  between the different orders in the coexistence region. While, the CDW phase is easily found inside the SC phase, where $T_{\mathrm{SC}}\ge T_{\mathrm{CDW}}$, the opposite is not true, i.e., in the region where $T_{\mathrm{CDW}}\ge T_{\mathrm{SC}}$,  SC is rapidly suppressed  when $|J_{d}|$ decreases, as shown in Fig.~\ref{fig:jxtvarufc}.
\begin{figure}[b]\centering
  \includegraphics[width=0.92\columnwidth]{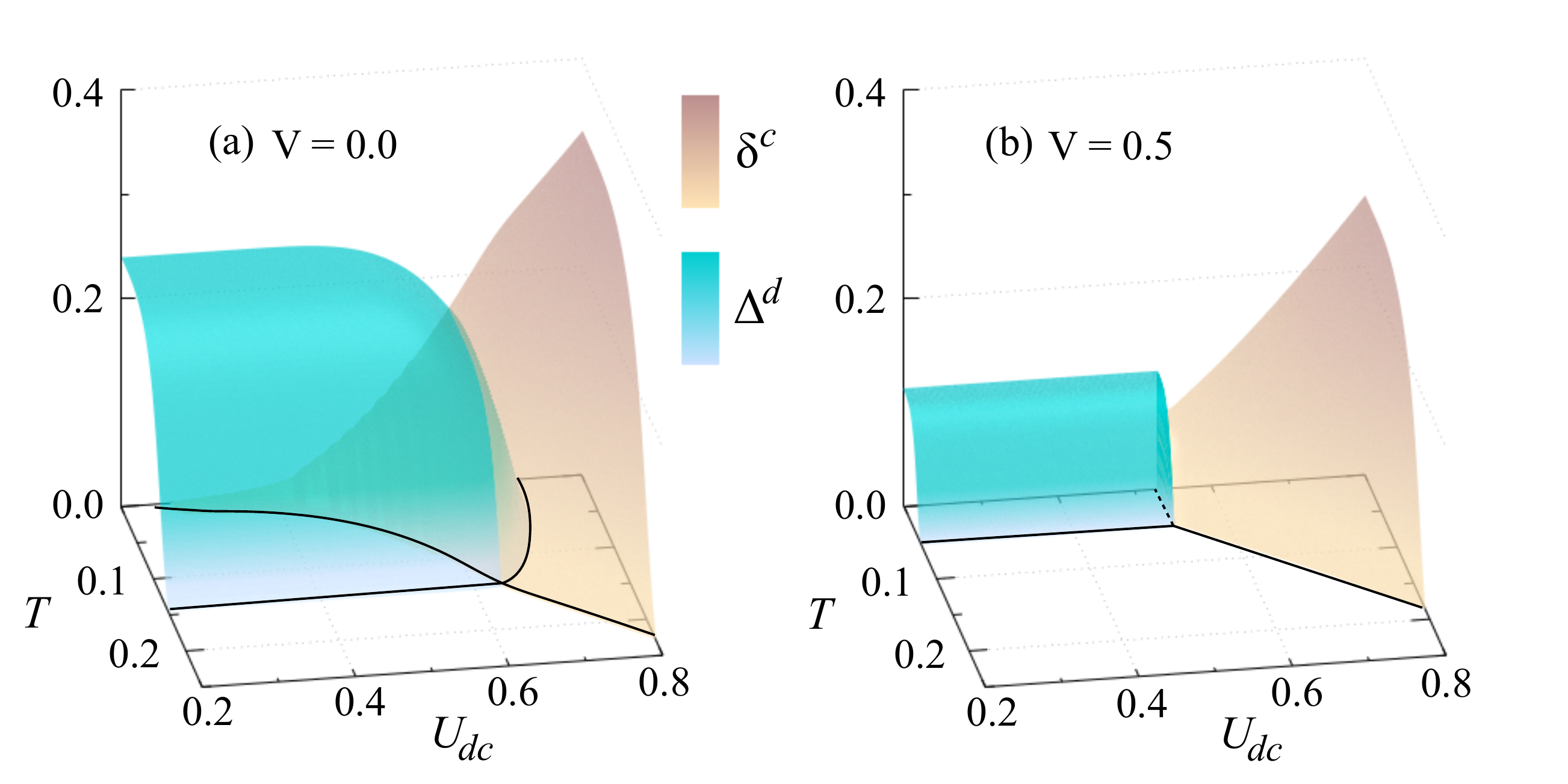}
  \caption{\label{fig:ufcxtvarv} Plots of $\delta^{c}$ and $\Delta^d$ as functions of $T$ and $U_{dc}$, for $J_{d}=-1.0$, $n_{\mathrm{tot}}=2.0$ and $\epsilon_{d0}=0.0$. (a) For $V=0.0$ and (b) $V=0.5$. For $V=0.0$ there is a small region of coexistence that is suppressed when we switch on $V$. The hybridization acts in detriment of both phases, but the SC order is destroyed faster when hybridization increases. Notice that the phase transitions are  second-order (continuous solid line) everywhere, except for $V=0.5$, where a first-order transition line (dashed line) separates the two ordered phases.} 
\end{figure}

This kind of behavior can also be observed as a function of the inter-band repulsion, as shown in Fig.~\ref{fig:ufcxtvarv}~(a). Again, the regime in which the CDW phase is inside the SC phase,   a coexistence region, is observed for small values of $U_{dc}$ ($U_{dc}\lesssim0.6$). However, for large values of $U_{dc}$, that favors the CDW state, superconductivity  hardly coexists. In Fig.~\ref{fig:ufcxtvarv}~(b) we show that hybridization is responsible for  destroying the region of coexistence. The magnitudes of order parameters are also affected, but SC is the most affected in this case. In the presence of hybridization the order parameters vanish abruptly  around $U_{dc}\approx 0.53$ at a first-order transition line (dashed line).

%%%%%%%%%%%%%%%%%%%%%%%%%%%%%%%%%%%%%%%%%%%%%%%%%%%%%%%%%%%%%%%%
\subsubsection{\texorpdfstring{$T$ {\it versus} V}{TxV}}

\begin{figure}[!h]\centering
  \includegraphics[width=0.92\columnwidth]{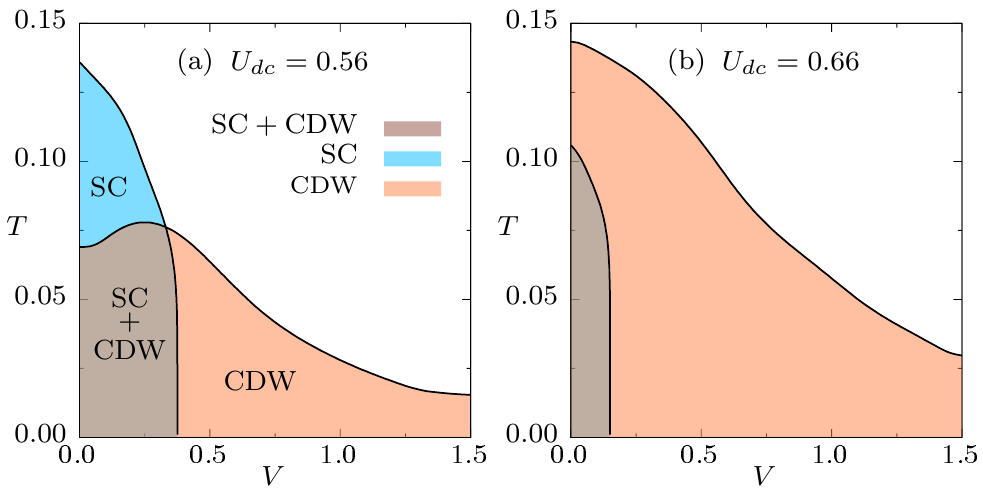}
  \caption{\label{fig:vixtempnt2} Critical temperatures (solid lines) as functions of $V$, for $J_{d}=-1.0$, $n_{\mathrm{tot}}=2.0$, and $\epsilon_{d0}=0.0$. (a) For $U_{dc}=0.56$ there is coexistence, with the CDW partially contained within the SC phase. In this case the highest critical temperature of the CDW phase occurs when superconductivity vanishes (b) For a small increase of $U_{dc}$ the CDW phase involves completely the SC phase.} 
\end{figure}

The ratio between hybridization and bandwidth, which  depends on the overlap of different wave functions and the depth of the $d$-band are both sensitive to pressure and chemical doping. Consequently, theoretical phase diagrams  obtained as a function of these quantities  have a direct relevance to those obtained experimentally when these parameters are varied~\cite{Reyes2019}. Here, we investigate the dependence of the critical temperatures (solid lines) on hybridization for different values of $U_{dc}$, as shown in Fig.~\ref{fig:vixtempnt2}.

Fig.~\ref{fig:vixtempnt2} shows a direct competition between the phases for two values of $U_{dc} < \left|J_d\right|$. We find coexistence for small hybridization $V$ for both values of $U_{dc}$. For the  smaller value, as shown in Fig.~\ref{fig:vixtempnt2}~(a),  SC is predominant for small $V$. For the larger value of $U_{dc}$,  Fig.~\ref{fig:vixtempnt2}~(b) shows that the CDW occupies a larger region in the phase diagram, specially at large $V$. The hybridization acts to destroy both phases, but SC goes faster to zero than CDW, even in the case where $T_{\mathrm{CDW}}<T_{\mathrm{SC}}$. All these transitions are second-order and are represented in the figures by continuous solid lines.

%%%%%%%%%%%%%%%%%%%%%%%%%%%%%%%%%%%%%%%%%%%%%%%%%%%%%%%%%%%%%%%%%%%%%%
\subsubsection{Effects of changing the relative depth of the bands}

Despite the discovery of many systems showing an interplay between CDW and SC to date, only a few of them show  coexistence of these phases at ambient pressure~\cite{Kumakura1996,Singh2005,Nagano2013,Luo2015}. In order to observe this behavior, it is necessary to apply external or chemical pressure in the system. It is also possible to introduce controlled disorder by, for example, electron irradiation or doping~\cite{Li2017,Cho2018}. In some cases these control parameters may give rise to quantum critical points where the CDW or SC phase vanishes at zero temperature.

In this section we consider a more subtle effect that consists in varying the relative depth of the band centers by changing the quantity $\epsilon_{d0}$.  In real systems we expect that this change in $\epsilon_{d0}$ can be implemented by considering systems where the elements responsible for the $d$-band of the material belong to different rows of the periodic table as $3d$, $4d$ or $5d$, but within the same column. As pointed out before, pressure can also alter the relative position of the bands.
\begin{figure}[t]
  \includegraphics[width=0.92\columnwidth]{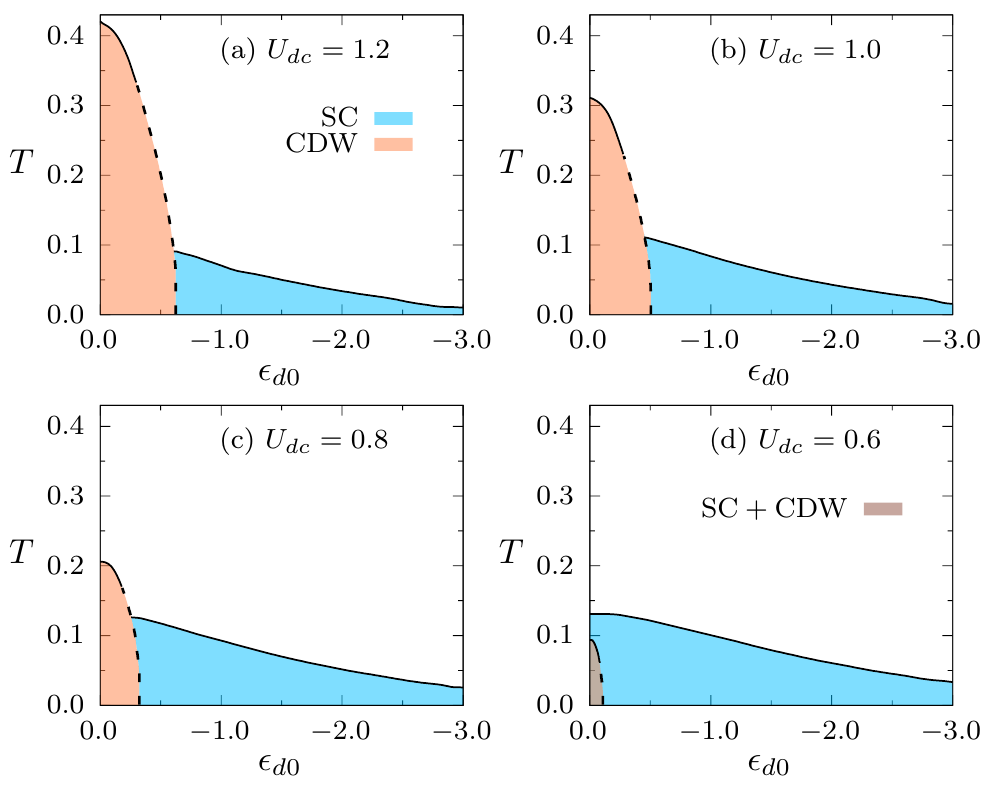}
  \caption{\label{fig:efxtvarufc} Critical temperatures (solid and dashed lines) as  functions of $\epsilon_{d0}$ for different values of $U_{dc}$, for $J_{d}=-1.0$, $n_{\mathrm{tot}}=2.0$ and $V=0.0$. Panels (a), (b) and (c) exhibit first (dashed line) or second-order  (solid line) transitions depending on the values of the parameters. (d) Coexistence occurs  only for small values of $U_{dc} =0.6$. As $|\epsilon_{d0}|$ increases, the CDW order is eventually suppressed for all values of $U_{dc}$, while the SC phase vanishes asymptotically. }
\end{figure}

Fig.~\ref{fig:efxtvarufc} depicts the results for the variation of the critical temperatures as functions of the relative depth of the bands, $\epsilon_{d0}$, and different values of $U_{dc}$. One can see that increasing of $|\epsilon_{d0}|$ is detrimental for both phases, but specially  for the CDW phase, which is suppressed for all values of $U_{dc}$ investigated. On the other hand, the SC phase vanishes for small values of $\left|\epsilon_{d0}\right|$ for $U_{dc} > 0.8$, but survives for large values of $\left|\epsilon_{d0}\right|$. In other words, $\epsilon_{d0}$ acts to shrink both phases, but the CDW order is more rapidly suppressed, while SC vanishes asymptotically.   Also note that there is coexistence of phases for $U_{dc}=0.6$, while for larger values of $U_{dc}$ we no longer observe it. Notice that, depending on the parameters, we obtain discontinuous first-order (dashed line) or second-order (continuous solid lines) phase transitions, as verified by calculations of the free energy density Eq.~\eqref{free_energy}.

%%%%%%%%%%%%%%%%%%%%%%%%%%%%%%%%%%%%%%%%%%%%%%%
\subsection{Away from half-filling}

In this section we explore the effects in the phase diagrams of changing the band occupations away from half-filling. This brings  new results and includes more realistic situations as  coexisting phases with a dominance of the CDW state for some regions of  parameter space.

%%%%%%%%%%%%%%%%%%%%%%%%%%%%%%%%%%%%%%%%%%%%%%%%%%%%%%%%%%%%%%%%%%%%%%%%%%%%%%%%
\subsubsection{\texorpdfstring{$T$ {\it versus} $n_{\mathrm{tot}}$}{T X ntot}}
\begin{figure}[b]
  \includegraphics[width=0.92\columnwidth]{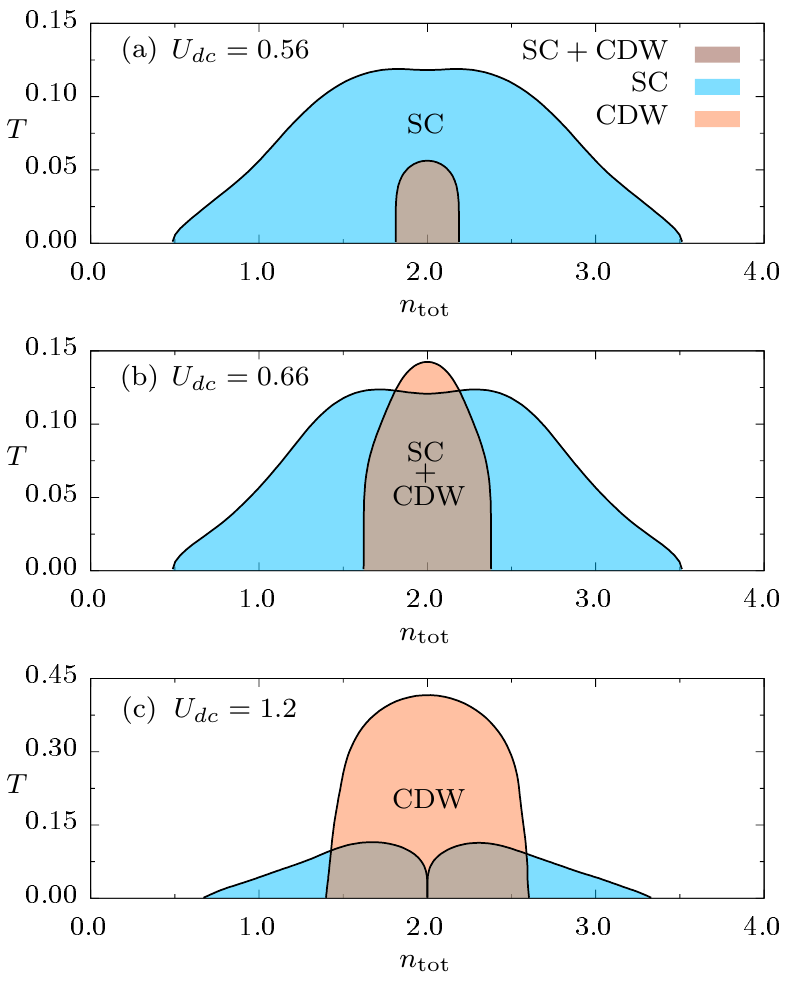}
  \caption{\label{fig:ntotxtemp} Critical temperatures as functions of $n_{\mathrm{tot}}$, for $J_{d}=-1.0$, $V=0.0$, and $\epsilon_{d0}=0.0$. (a) For small values of $U_{dc}$, CDW emerges around half-filling values. In (b) and (c), as we increase the magnitude of $U_{dc}$, the CDW phase spreads out from $n_{\mathrm{tot}} = 2.0$. Notice in (c)  that the SC phase is suppressed at half-filling  for large $U_{dc}$.} 
\end{figure}
The critical temperatures as functions of the total number of particles are shown in Fig.~\ref{fig:ntotxtemp}. The CDW phase appears at half-filling and spreads out as we increase $U_{dc}$. Concomitantly, the region of the phase diagram occupied by the SC phase is reduced. Moreover, for sufficiently large $U_{dc}$,  SC vanishes at exactly half-filling.

Also notice, from Fig.~\ref{fig:ntotxtemp} that both critical temperatures are symmetric around $n_{\mathrm{tot}}=2.0$, and this holds even when $V \neq 0$ (not shown). However, for finite, but small values of $\left|\epsilon_{d0}\right|$, this symmetry is lost and both critical regions are shifted towards $n_{\mathrm{tot}}< 2.0$. Increasing further $\left|\epsilon_{d0}\right|$, for instance for $\epsilon_{d0} = - 1.0$ and $U_{dc}=1.2$ as in Fig.~\ref{fig:ntotxtemp}~(c), the SC lobes merge and there is a finite superconducting critical temperature at half-filling. In addition, the region of superconductivity in the phase diagram  is slightly shifted towards $n_{\mathrm{tot}}< 2.0$. On the other hand, for this value of $\epsilon_{d0}$ and $U_{dc}=1.2$ (not shown), the  CDW phase has been completely washed out from the phase diagram,   corroborating the deleterious effect of the band shift on this phase.

%%%%%%%%%%%%%%%%%%%%%%%%%%%%%%%%%%%%%%%%%%
\subsubsection{\texorpdfstring{$T$ {\it versus} $V$}{TxV}}

\begin{figure}[H]\centering
  \includegraphics[width=0.92\columnwidth]{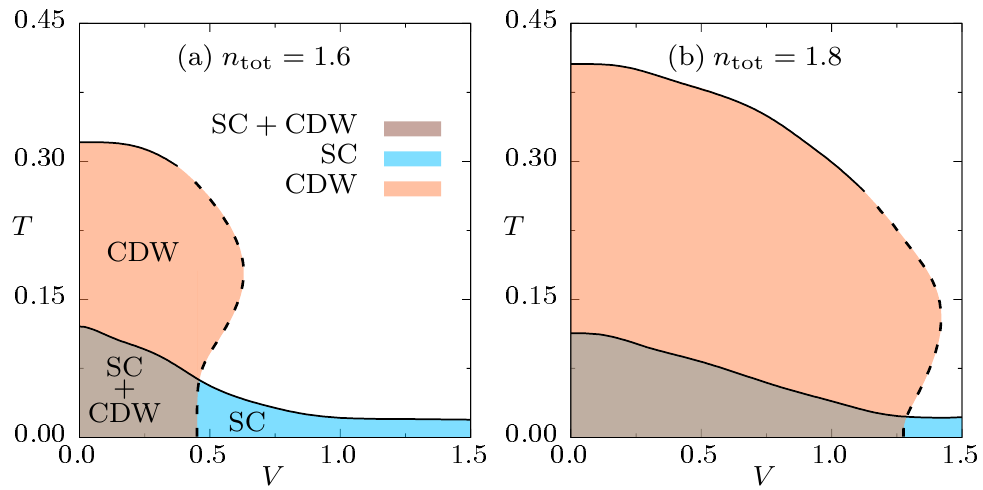}
  \caption{\label{fig:vixtemp} Critical temperatures as a function of $V$, for $U_{dc}=1.2 > \left|J_d\right|$, $J_{d}=-1.0$, and $\epsilon_{d0}=0.0$. (a) $n_{\mathrm{tot}}=1.6$ and (b) $n_{\mathrm{tot}}=1.8$. In both cases the phase diagrams present a reentrant behavior that is  characteristic of  first-order (dashed line) phase transitions. Also note the asymptotic behavior for SC as a function of V in this case.}
\end{figure}

Figs.~\ref{fig:vixtemp}~(a) e~(b) exhibit a qualitative agreement with experimental results for  compounds that present a discontinuous vanishing of the CDW transition with pressure~\cite{Shen2020}, although with a persistent SC phase at low temperatures. The reentrant behavior of $T_{\mathrm{CDW}}$ for large values of $U_{dc}$ and away from half-filling  is a signature of a first-order phase transition~\cite{Khomskii2010}. It is worth to emphasize the asymptotic behavior for SC as a function of $V$ in this case. This is very different when compared to the half-filling results, see Fig.~\ref{fig:vixtempnt2}.

%%%%%%%%%%%%%%%%%%%%%%%%%%%%%%%%%%%%%%%%%%%%%%%%%%%%%%%%%%%%%%%%%%%%%%%%%%%%%
\subsubsection{Effects of changing the relative depth of the bands for \texorpdfstring{$n_{\mathrm{tot}} \ne 2.0$}{ntot =| 2}}
\begin{figure}[!h]\centering 
  \includegraphics[width=0.92\columnwidth]{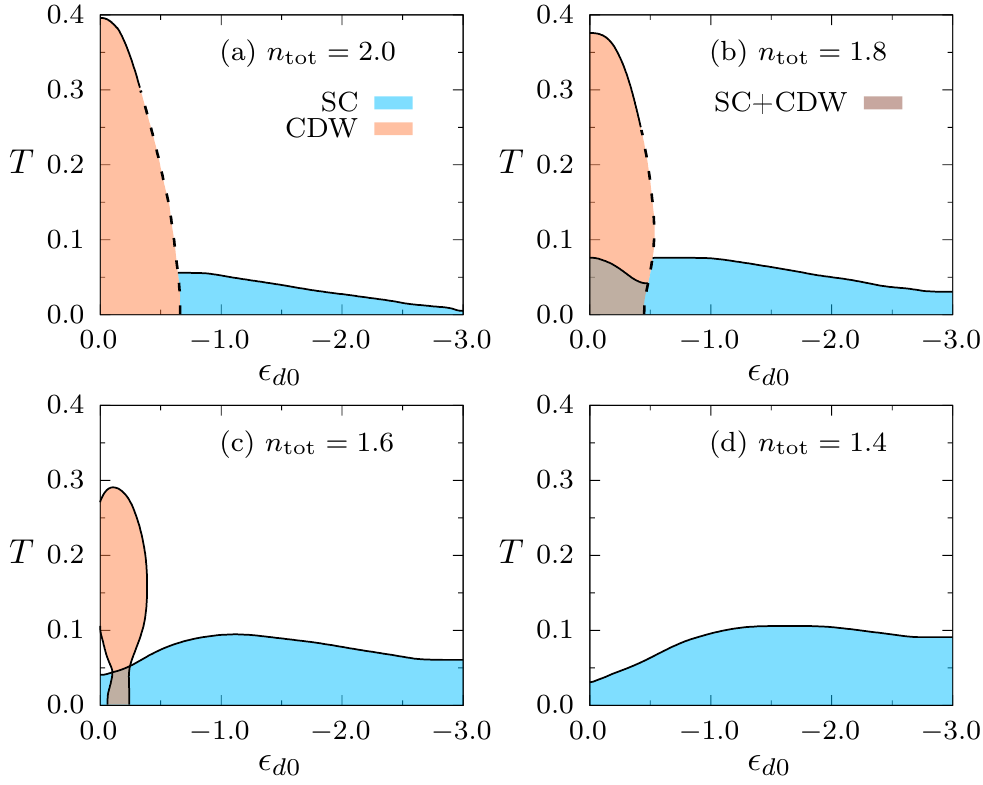}
  \caption{\label{fig:ef0xtvarnt} Critical temperatures as functions of $\epsilon_{d0}$ for different values of $n_{\mathrm{tot}}$, for $J_{d}=-1.0$, $U_{dc}=1.2$ and $V=0.5$. (a) At half-filling there is no coexistence. However, a small region of coexistence appears for (b) $n_{\mathrm{tot}} = 1.8$ and (c) $n_{\mathrm{tot}} = 1.6$. (d) For $n_{\mathrm{tot}} = 1.4$ appears SC order only.}
\end{figure}

In order to investigate the combined effect of the total number of particles $n_{\mathrm{tot}} \neq 2.0$ and $\epsilon_{d0}$ in the phase diagrams, we present in Fig.~\ref{fig:ef0xtvarnt} our results for the variation of the critical temperatures as functions of $\epsilon_{d0}$ and $n_{\mathrm{tot}}$. For half-filling and $U_{dc}=1.2$ there is no coexistence region, as already reported in an earlier section, see Fig.~\ref{fig:efxtvarufc}. However, for $n_{\mathrm{tot}}=1.8$ and $n_{\mathrm{tot}}=1.6$ the system presents  a coexistence region for small values of $\left|\epsilon_{d0}\right|$. For $n_{\mathrm{tot}}=1.4$, only SC order exists.

Summing up, the phase diagram of the system is sensitive to the variation of  $\epsilon_{d0}$. The general effect of increasing $|\epsilon_{d0}|$ in all scenarios we studied is to destroy faster the CDW phase, while the SC phase persists even for large values of $|\epsilon_{d0}|$. 

\begin{figure}[b]\centering
  \includegraphics[width=0.92\columnwidth]{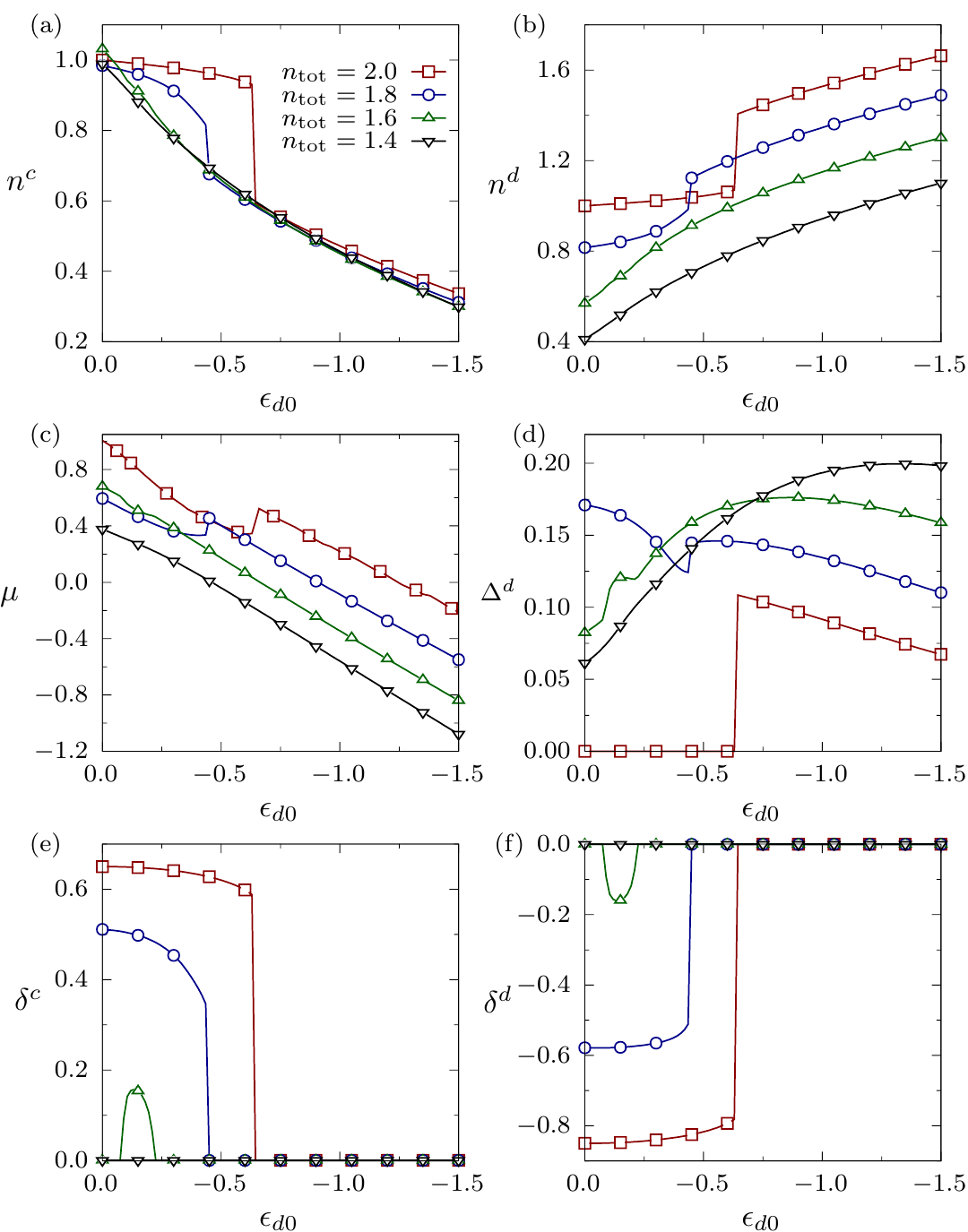}
  \caption{\label{fig:parametersxef0_t001} Occupation numbers (a) $n^c$ and (b) $n^d$,   chemical potential $\mu$ (c), and order parameters (d) $\Delta^{d}$, (e) $\delta^{c}$ and (f) $\delta^{d}$ as functions of $\epsilon_{d0}$ for $T\rightarrow0$, and $J_d=-1.0$, $U_{dc}=1.2$, and $V=0.5$. For $n_{\mathrm{tot}}=2.0$ and $\epsilon_{d0}=0.0$, we have $n^c=n^d$ and the CDW order parameters attain their maximum values, but there is no SC phase. As $|\epsilon_{d0}|$ increases, the occupations numbers become different. They change abruptly for $|\epsilon_{d0}|\sim 0.65$ when SC appears and the CDW order vanishes. For $n_{\mathrm{tot}} \ne 2.0$ these behaviors become smooth although preserving the same tendencies. }
\end{figure}
The persistence of $\Delta^d$ and the fast destruction of the CDW phase as a function of $\epsilon_{d0}$ can be rationalized by the variation of $n^c$, $n^d$, $\mu$, $\Delta^{d}$, $\delta^{c}$ and $\delta^{d}$ as a function of $\epsilon_{d0}$ shown in Fig.~\ref{fig:parametersxef0_t001}. Keeping $n_{\mathrm{tot}}$ fixed, there is an imbalance between $n^c$ and $n^d$ when $|\epsilon_{d0}|$ increases, which disfavors the CDW phase, as can be seen in Fig.~\ref{fig:parametersxef0_t001}. For fixed $n_{\mathrm{tot}}=2.0$, $n^c=n^d$  when $\epsilon_{d0}=0$. In this case the CDW order parameters, $\delta^{c(d)} \neq 0$ attain their maximum values while the SC order parameter vanishes. As $|\epsilon_{d0}|$ increases  there is an abrupt variation of the occupation numbers, while the chemical potential $\mu$ always decreases, and the CDW phase is destroyed given place to the SC phase.

When the occupation deviates from half-filling, the amplitude  of the CDW order parameter decreases, allowing for an initial coexistence of phases that disappears as $\delta n=|n_{\mathrm{tot}}-2.0|$ increases, as also shown in Fig.~\ref{fig:ntotxtemp}~(c). The finite hybridization included here, $V=0.5$, is responsible for a reentrant behavior that is seen for $n_{\mathrm{tot}}=1.6$.

%%%%%%%%%%%%%%%%%%%%%%%%%%%%%%%%%%%%%%%%%%%%%%%%%%%%%%%%%%%%%%%%%%%%%%%%%
\subsection{The inhomogeneous superconductivity}\label{inhomogeneousSC}

A possible ground state of a SC is a pair density wave phase where the superconducting order parameter varies in space according to some wave-vector $\vQ$~\cite{Chikina2020}. In general, this Fulde–Ferrell–Larkin–Ovchinnikov~(FFLO)~\cite{Fulde1964,Larkin1965} or pair density wave state~\cite{Agterberg2020} is expected to occur in the presence of a sufficiently strong magnetic field.  Here we raise the question, whether in the presence of charge modulation as occurs in a CDW state this pair density wave state can be energetically favored. We consider a modulated SC ground state characterized by the same wave-vector of the CDW  ($\pi,\pi$) and investigate the presence of these phases in the phase diagram. We find that they strongly compete and there is no coexistence in the phase diagrams.

\begin{figure}[!h]\centering 
  \includegraphics[width=0.92\columnwidth]{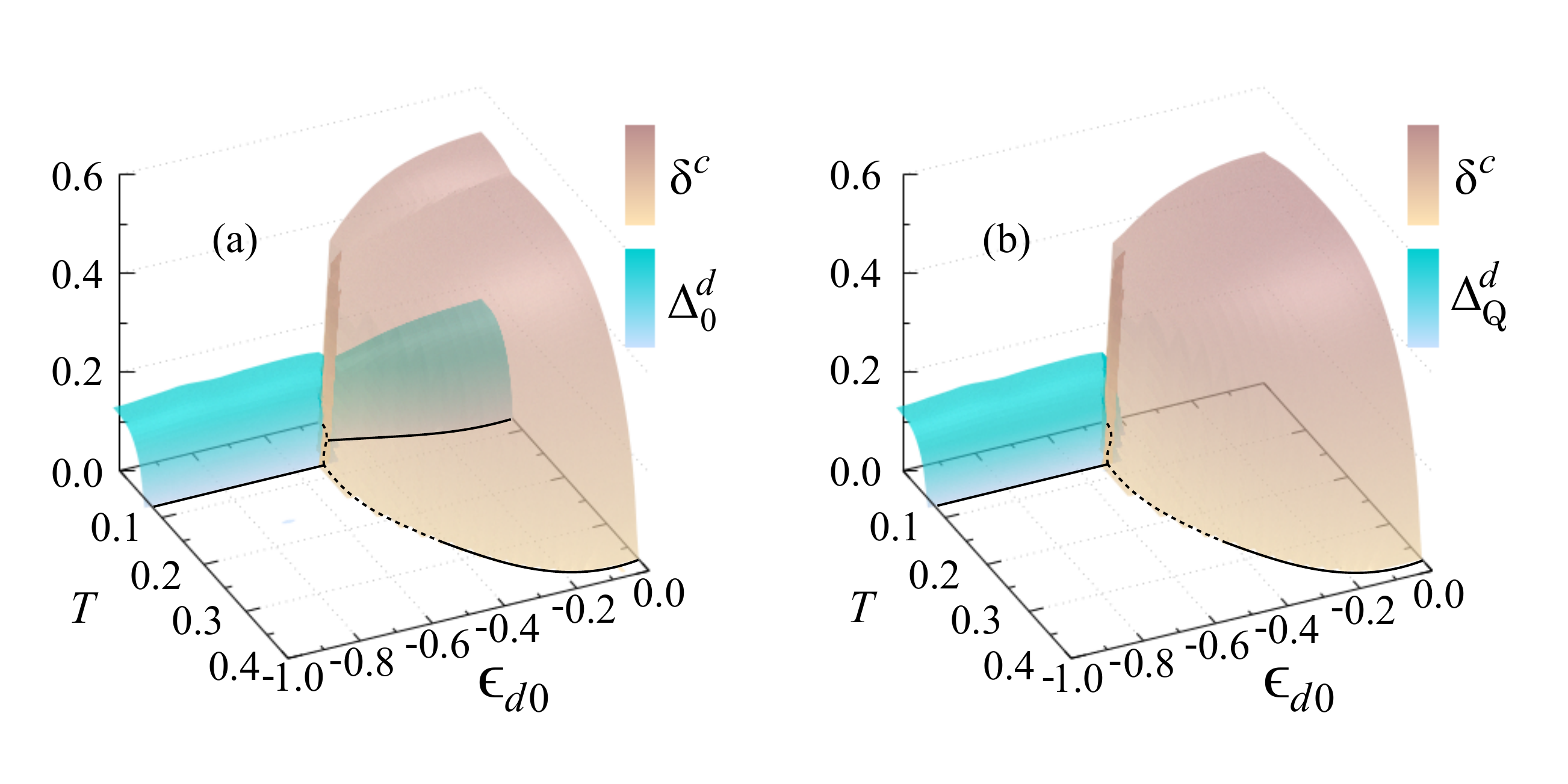}
  \caption{\label{fig:ef0xt_nt18compara} Plots of $\delta^c$ and $\Delta^d$ as functions of $T$ and $\epsilon_{d0}$. (a) For homogeneous SC ($\Delta^{d}_{0}$) and (b) inhomogeneous SC ($\Delta^{d}_{\vQ}$), i.e., a pair density wave SC state. In the latter case there is no coexistence between SC and CDW phases.  We used  $J_{d}=-1.0$, $U_{dc}=1.2$, $V=0.5$ and $n_{\mathrm{tot}}=1.8$. }
\end{figure}

In Fig.~{\ref{fig:ef0xt_nt18compara}} we compare the phase diagrams for CDW phase with homogeneous ($\Delta^{d}=\Delta^{d}_{0}$) and inhomogeneous ($\Delta^{d}_{\vQ}$) SC for $n_{\mathrm{tot}}=1.8$. One can see that $\Delta^{d}_{\vQ}$ suppress the coexistence region that appears in the $\Delta^{d}_{0}$ case.

%%%%%%%%%%%%%%%%%%%%%%%%%%%%%%%%%%%%%%%%%%%%%%%%%%%%
\subsection{Coexistence and first order transitions}
 
\begin{figure}[b]\centering 
  \includegraphics[width=0.92\columnwidth]{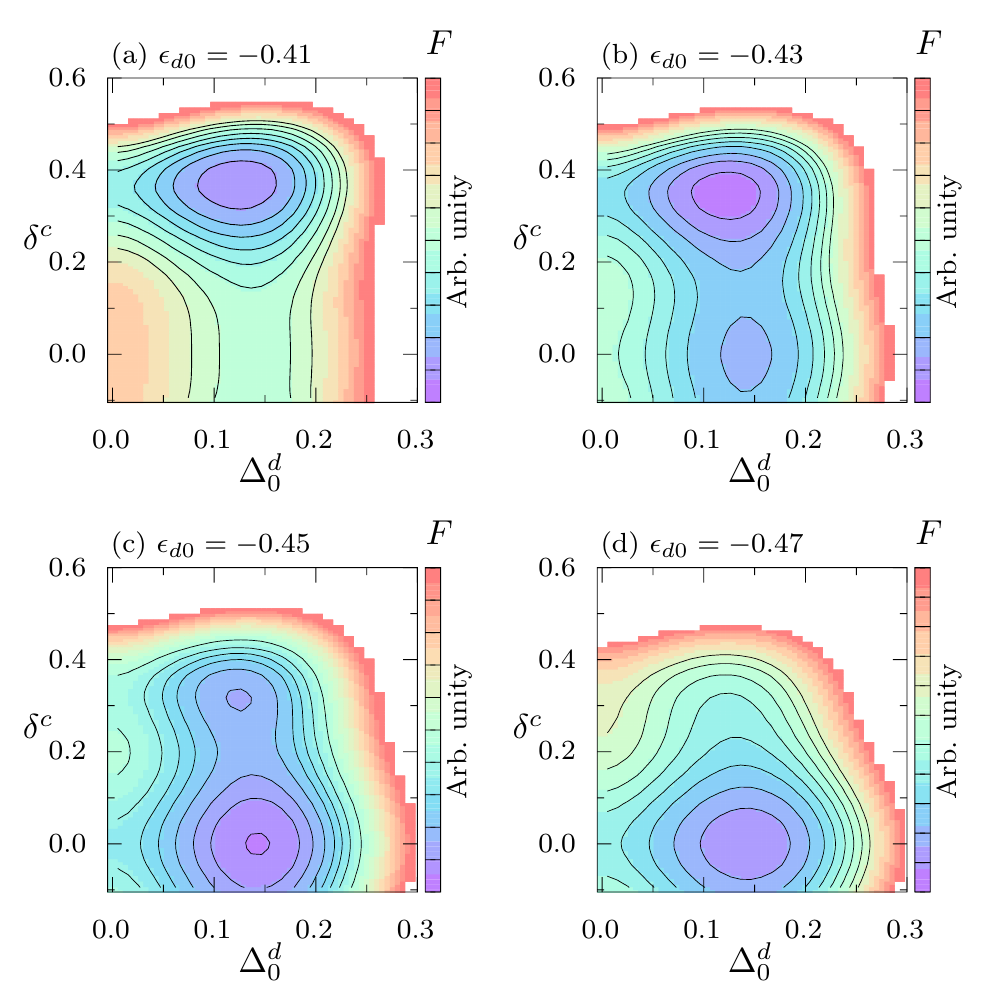}
  \caption{\label{fig:fener_dscxdc} Free energy density map for $\delta^c$ and  $\Delta^{d}_{0}$ for $n_{\mathrm{tot}}=1.8$, $T=0.001$, $J_d=-1.0$, $U_{dc}=1.2$, and $V=0.5$. The four panels correspond to different values of $\epsilon_{d0}$. (a) For $\epsilon_{d0}=-0.41$  the system is in a homogeneous  phase, with both order parameters finite and only one minimum is observed. (b) For $\epsilon_{d0}=-0.43$ there are two minima with a metastable one at $\delta^c=0$, $\Delta^{d}_{0}$ finite. In (c), for $\epsilon_{d0}=-0.45$, these minima have exchange stability with an intervening  first-order transition.  (d) Finally, for $\epsilon_{d0}=-0.47$ there is only minimum at $\delta^c=0$ and $\Delta^{d}_{0} \neq 0$ that corresponds to a pure SC phase.}
\end{figure}

In order to investigate the nature of the coexistence of phases, as shown in Fig.~{\ref{fig:ef0xt_nt18compara}}~(a), which was found in several cases in this study, it is necessary to obtain  the free energy density of the system as a function of the order parameters. These calculations are costly numerically and although they were performed to determine the nature of different transitions,  we are going to discuss in detail just the phase diagram shown in Fig.~\ref{fig:ef0xt_nt18compara}~(a). In this case, coexistence occurs as a function of the shift between the band centers. Fig.~\ref{fig:fener_dscxdc} shows the free energy density as a function of the order parameters for the same  values of the parameters of Fig.~\ref{fig:ef0xt_nt18compara}~(a).
For  $n_{\mathrm{tot}}= 1.8$ and $\epsilon_{d0}=-0.41$ we observe a single minimum in the free energy at finite values of both $\delta^c$ and $\Delta_0^d$, as can be seen in Fig.~\ref{fig:fener_dscxdc}~(a). Then, for $\epsilon_{d0}=-0.41$, the system presents homogeneous coexistence between the CDW and SC phases.  As $\left|\epsilon_{d0}\right|$ increases  a new  minimum appears at $\delta^{c}=0$, but at a finite value of the superconducting order parameter, $\Delta^{d}_{0}$, as shown in Fig.~\ref{fig:fener_dscxdc}~(b). This is a metastable minimum as it has a higher energy than the minimum with both order parameters finite. Further increasing $\left|\epsilon_{d0}\right|$, at $\epsilon_{d0}=-0.45$ these minima have exchanged stability, as shown in Fig.~\ref{fig:fener_dscxdc}~(c). Eventually the minimum corresponding to both order parameters finite disappears, as one goes deep in the pure superconducting phase leaving only the SC minimum as in Fig.~\ref{fig:fener_dscxdc}~(d) for $\epsilon_{d0}=-0.47$.

The phases that have exchanged stability close to $\epsilon_{d0}= -0.45$ are, a phase with homogeneous coexistence of SC and CDW and a phase that is purely superconducting. Since there  is an intervening first-order transition, we may expect that close to it the system presents regions with both CDW and SC orders coexisting with regions that are  purely superconducting.

%%%%%%%%%%%%%%%%%%%%%%%%%%%%%%%%%%%%%%%%%%%%%%%%%%%%%%%
\section{Conclusions}\label{conclusions}

Electronic interactions on correlated systems  may give rise, at low temperatures, to different ground states as external control parameters, such as pressure, doping or magnetic field are varied~\cite{Nei2020}. In these systems, competition or coexistence of different phases described using multiple order parameters might be observed experimentally. In particular, the coexistence/competition between SC and CDW phases has been recently reported in several compounds~\cite{Morosan2006,Kusmartseva2009,Gruner2017}. In some cases, but not always, CDW seems to be accompanied by a first-order structural phase transition~\cite{Gruner2017,Shen2020}. It is also generally observed that pressure suppresses the CDW order but enhances the superconducting transition temperature $T_{\mathrm{SC}}$~\cite{Shen2020}.

In order to clarify the role of the relevant interactions in systems exhibiting SC and CDW orderings we carried out in this paper the study of a two-band model taking into account the interplay between these phases in a square lattice. The choice for this lattice is that it is computationally simpler, but still expected to describe results in three-dimensions within the mean-field treatment of the interactions that we use. Our main motivation are inter-metallic systems that present a $d$-band of moderately correlated electrons coexisting with a large $sp$-band. The CDW phase emerges from inter-band Coulomb interactions and SC is due to an intra-band attractive interaction.  The moderate strength of the correlations in the materials we are interested justifies our mean-field approach~\cite{Brydon2005,Farka2008,Banerjee2018} that makes the problem numerically tractable. Furthermore, we neglected the $\vk$-dependence of the hybridization between the bands and of the interactions. For completeness, we also studied the possible coexistence of CDW with both, a homogeneous SC and that with a pair density wave with the same modulation wave-vector of the CDW.

The many-body problem is approached using a HF mean-field decoupling that has been widely applied to deal with SC and CDW in different models~\cite{Brydon2005,Farka2008,Banerjee2018}. We proceed writing the Hamiltonian in the Nambu representation. The eight by eight Hamiltonian matrix is diagonalized numerically. We obtain the eigenvalues and eigenvectors numerically and minimize the free energy density with respect to different parameters to obtain the phase diagrams as functions of external control parameters, such as strength of the interactions, hybridization, total number of particles and relative depth of the bands.

We show that for the model studied there is a direct competition between CDW and SC orders, but depending on the parameters these phases might coexist. We find that the CDW phase emerges around half-filling, in agreement with other approaches~\cite{Brydon2005,Farka2008}, and spreads out over the phase diagram as the inter-band Coulomb interaction increases. Away from half-filling, we obtain a qualitative agreement with reported results for compounds that exhibit a discontinuous vanishing of the CDW transition~\cite{Shen2020}. Moreover, we find a reentrant behavior of $T_{\mathrm{CDW}}$ for large values of $U_{dc}$, which is an intrinsic signature of a first-order phase transition~\cite{Khomskii2010}.

We also obtained that as the distance between the band centers increases, the CDW phase vanishes and only SC survives. The latter is robust to a change of the relative depth of the bands. In other words, as we increase $\left|\epsilon_{d0}\right|$ the coexistence of phases is suppressed, remaining the SC phase only. On the other hand, the pair density wave SC state does not support coexistence with the CDW.

Our results reproduce most of the experimental observations in CDW-SC inter-metallic compounds with d-bands and dimensions $d \ge 2$. It contributes to identify the relevant interactions and parameters that determine the behavior of these systems. It throws light on the nature of the coexistence between CDW-SC phases and shows that in some cases these phases may coexist homogeneously. We found both continuous and first-order transitions depending on the parameters of the model. The phase diagrams are very sensitive to the specific values chosen for the parameters and for this reason we did not compare our results with any particular system. The connection of our results with experiments arises since both the hybridization and relative depth of the bands are pressure dependent. Besides these compounds can be doped with different elements and this is accounted in the model by  varying the total number of electrons.

%%%%%%%%%%%%%%%%%%%%%%%%%%%%%%%%%%%%%%%%%%
\section{Acknowledgments}

M.A.C. would like to thank the Brazilian agencies FAPERJ, CAPES, and CNPq for partial financial support. N.L. would like to thank the FAPERJ for the post doctoral fellowship of the \textit{Programa de Pós-Doutorado Nota 10 - 2020}. D.R. would like to thank the Brazilian Center for Research in Physics (CBPF) where part of this work was performed. Finally, we would like to thank the COTEC (CBPF) since the numerical calculations were performed on the \textit{Cluster HPC}.

%%%%%%%%%%%%%%%%%%%%%%%%%%%%%%%%%%%%%%%%
\bibliographystyle{apsrev4-2}
\bibliography{biblio_cdwsc_201103}

\begin{thebibliography}{83}%
\makeatletter
\providecommand \@ifxundefined [1]{%
 \@ifx{#1\undefined}
}%
\providecommand \@ifnum [1]{%
 \ifnum #1\expandafter \@firstoftwo
 \else \expandafter \@secondoftwo
 \fi
}%
\providecommand \@ifx [1]{%
 \ifx #1\expandafter \@firstoftwo
 \else \expandafter \@secondoftwo
 \fi
}%
\providecommand \natexlab [1]{#1}%
\providecommand \enquote  [1]{``#1''}%
\providecommand \bibnamefont  [1]{#1}%
\providecommand \bibfnamefont [1]{#1}%
\providecommand \citenamefont [1]{#1}%
\providecommand \href@noop [0]{\@secondoftwo}%
\providecommand \href [0]{\begingroup \@sanitize@url \@href}%
\providecommand \@href[1]{\@@startlink{#1}\@@href}%
\providecommand \@@href[1]{\endgroup#1\@@endlink}%
\providecommand \@sanitize@url [0]{\catcode `\\12\catcode `\$12\catcode
  `\&12\catcode `\#12\catcode `\^12\catcode `\_12\catcode `\%12\relax}%
\providecommand \@@startlink[1]{}%
\providecommand \@@endlink[0]{}%
\providecommand \url  [0]{\begingroup\@sanitize@url \@url }%
\providecommand \@url [1]{\endgroup\@href {#1}{\urlprefix }}%
\providecommand \urlprefix  [0]{URL }%
\providecommand \Eprint [0]{\href }%
\providecommand \doibase [0]{https://doi.org/}%
\providecommand \selectlanguage [0]{\@gobble}%
\providecommand \bibinfo  [0]{\@secondoftwo}%
\providecommand \bibfield  [0]{\@secondoftwo}%
\providecommand \translation [1]{[#1]}%
\providecommand \BibitemOpen [0]{}%
\providecommand \bibitemStop [0]{}%
\providecommand \bibitemNoStop [0]{.\EOS\space}%
\providecommand \EOS [0]{\spacefactor3000\relax}%
\providecommand \BibitemShut  [1]{\csname bibitem#1\endcsname}%
\let\auto@bib@innerbib\@empty
%</preamble>
\bibitem [{\citenamefont {Morosan}\ \emph {et~al.}(2006)\citenamefont
  {Morosan}, \citenamefont {Zandbergen}, \citenamefont {Dennis}, \citenamefont
  {Bos}, \citenamefont {Onose}, \citenamefont {Klimczuk}, \citenamefont
  {Ramirez}, \citenamefont {Ong},\ and\ \citenamefont {Cava}}]{Morosan2006}%
  \BibitemOpen
  \bibfield  {author} {\bibinfo {author} {\bibfnamefont {E.}~\bibnamefont
  {Morosan}}, \bibinfo {author} {\bibfnamefont {H.~W.}\ \bibnamefont
  {Zandbergen}}, \bibinfo {author} {\bibfnamefont {B.~S.}\ \bibnamefont
  {Dennis}}, \bibinfo {author} {\bibfnamefont {J.~W.~G.}\ \bibnamefont {Bos}},
  \bibinfo {author} {\bibfnamefont {Y.}~\bibnamefont {Onose}}, \bibinfo
  {author} {\bibfnamefont {T.}~\bibnamefont {Klimczuk}}, \bibinfo {author}
  {\bibfnamefont {A.~P.}\ \bibnamefont {Ramirez}}, \bibinfo {author}
  {\bibfnamefont {N.~P.}\ \bibnamefont {Ong}},\ and\ \bibinfo {author}
  {\bibfnamefont {R.~J.}\ \bibnamefont {Cava}},\ }\href
  {https://doi.org/10.1038/nphys360} {\bibfield  {journal} {\bibinfo  {journal}
  {Nature Physics}\ }\textbf {\bibinfo {volume} {2}},\ \bibinfo {pages} {544}
  (\bibinfo {year} {2006})}\BibitemShut {NoStop}%
\bibitem [{\citenamefont {Kusmartseva}\ \emph {et~al.}(2009)\citenamefont
  {Kusmartseva}, \citenamefont {Sipos}, \citenamefont {Berger}, \citenamefont
  {Forr{\'{o}}},\ and\ \citenamefont {Tuti{\v{s}}}}]{Kusmartseva2009}%
  \BibitemOpen
  \bibfield  {author} {\bibinfo {author} {\bibfnamefont {A.~F.}\ \bibnamefont
  {Kusmartseva}}, \bibinfo {author} {\bibfnamefont {B.}~\bibnamefont {Sipos}},
  \bibinfo {author} {\bibfnamefont {H.}~\bibnamefont {Berger}}, \bibinfo
  {author} {\bibfnamefont {L.}~\bibnamefont {Forr{\'{o}}}},\ and\ \bibinfo
  {author} {\bibfnamefont {E.}~\bibnamefont {Tuti{\v{s}}}},\ }\bibfield
  {journal} {\bibinfo  {journal} {Physical Review Letters}\ }\textbf {\bibinfo
  {volume} {103}},\ \href {https://doi.org/10.1103/physrevlett.103.236401}
  {10.1103/physrevlett.103.236401} (\bibinfo {year} {2009})\BibitemShut
  {NoStop}%
\bibitem [{\citenamefont {Gruner}\ \emph {et~al.}(2017)\citenamefont {Gruner},
  \citenamefont {Jang}, \citenamefont {Huesges}, \citenamefont {Cardoso-Gil},
  \citenamefont {Fecher}, \citenamefont {Koza}, \citenamefont {Stockert},
  \citenamefont {Mackenzie}, \citenamefont {Brando},\ and\ \citenamefont
  {Geibel}}]{Gruner2017}%
  \BibitemOpen
  \bibfield  {author} {\bibinfo {author} {\bibfnamefont {T.}~\bibnamefont
  {Gruner}}, \bibinfo {author} {\bibfnamefont {D.}~\bibnamefont {Jang}},
  \bibinfo {author} {\bibfnamefont {Z.}~\bibnamefont {Huesges}}, \bibinfo
  {author} {\bibfnamefont {R.}~\bibnamefont {Cardoso-Gil}}, \bibinfo {author}
  {\bibfnamefont {G.~H.}\ \bibnamefont {Fecher}}, \bibinfo {author}
  {\bibfnamefont {M.~M.}\ \bibnamefont {Koza}}, \bibinfo {author}
  {\bibfnamefont {O.}~\bibnamefont {Stockert}}, \bibinfo {author}
  {\bibfnamefont {A.~P.}\ \bibnamefont {Mackenzie}}, \bibinfo {author}
  {\bibfnamefont {M.}~\bibnamefont {Brando}},\ and\ \bibinfo {author}
  {\bibfnamefont {C.}~\bibnamefont {Geibel}},\ }\href
  {https://doi.org/10.1038/nphys4191} {\bibfield  {journal} {\bibinfo
  {journal} {Nature Physics}\ }\textbf {\bibinfo {volume} {13}},\ \bibinfo
  {pages} {967} (\bibinfo {year} {2017})}\BibitemShut {NoStop}%
\bibitem [{\citenamefont {Zhao}\ \emph {et~al.}(2007)\citenamefont {Zhao},
  \citenamefont {Ou}, \citenamefont {Wu}, \citenamefont {Xie}, \citenamefont
  {Zhang}, \citenamefont {Shen}, \citenamefont {Wei}, \citenamefont {Yang},
  \citenamefont {Dong}, \citenamefont {Arita}, \citenamefont {Namatame},
  \citenamefont {Taniguchi}, \citenamefont {Chen},\ and\ \citenamefont
  {Feng}}]{Zhao2007}%
  \BibitemOpen
  \bibfield  {author} {\bibinfo {author} {\bibfnamefont {J.~F.}\ \bibnamefont
  {Zhao}}, \bibinfo {author} {\bibfnamefont {H.~W.}\ \bibnamefont {Ou}},
  \bibinfo {author} {\bibfnamefont {G.}~\bibnamefont {Wu}}, \bibinfo {author}
  {\bibfnamefont {B.~P.}\ \bibnamefont {Xie}}, \bibinfo {author} {\bibfnamefont
  {Y.}~\bibnamefont {Zhang}}, \bibinfo {author} {\bibfnamefont {D.~W.}\
  \bibnamefont {Shen}}, \bibinfo {author} {\bibfnamefont {J.}~\bibnamefont
  {Wei}}, \bibinfo {author} {\bibfnamefont {L.~X.}\ \bibnamefont {Yang}},
  \bibinfo {author} {\bibfnamefont {J.~K.}\ \bibnamefont {Dong}}, \bibinfo
  {author} {\bibfnamefont {M.}~\bibnamefont {Arita}}, \bibinfo {author}
  {\bibfnamefont {H.}~\bibnamefont {Namatame}}, \bibinfo {author}
  {\bibfnamefont {M.}~\bibnamefont {Taniguchi}}, \bibinfo {author}
  {\bibfnamefont {X.~H.}\ \bibnamefont {Chen}},\ and\ \bibinfo {author}
  {\bibfnamefont {D.~L.}\ \bibnamefont {Feng}},\ }\bibfield  {journal}
  {\bibinfo  {journal} {Physical Review Letters}\ }\textbf {\bibinfo {volume}
  {99}},\ \href {https://doi.org/10.1103/physrevlett.99.146401}
  {10.1103/physrevlett.99.146401} (\bibinfo {year} {2007})\BibitemShut
  {NoStop}%
\bibitem [{\citenamefont {Wilson}\ \emph {et~al.}(1975)\citenamefont {Wilson},
  \citenamefont {Salvo},\ and\ \citenamefont {Mahajan}}]{Wilson1975}%
  \BibitemOpen
  \bibfield  {author} {\bibinfo {author} {\bibfnamefont {J.}~\bibnamefont
  {Wilson}}, \bibinfo {author} {\bibfnamefont {F.~D.}\ \bibnamefont {Salvo}},\
  and\ \bibinfo {author} {\bibfnamefont {S.}~\bibnamefont {Mahajan}},\ }\href
  {https://doi.org/10.1080/00018737500101391} {\bibfield  {journal} {\bibinfo
  {journal} {Advances in Physics}\ }\textbf {\bibinfo {volume} {24}},\ \bibinfo
  {pages} {117} (\bibinfo {year} {1975})}\BibitemShut {NoStop}%
\bibitem [{\citenamefont {Sipos}\ \emph {et~al.}(2008)\citenamefont {Sipos},
  \citenamefont {Kusmartseva}, \citenamefont {Akrap}, \citenamefont {Berger},
  \citenamefont {Forr{\'{o}}},\ and\ \citenamefont {Tuti{\v{s}}}}]{Sipos2008}%
  \BibitemOpen
  \bibfield  {author} {\bibinfo {author} {\bibfnamefont {B.}~\bibnamefont
  {Sipos}}, \bibinfo {author} {\bibfnamefont {A.~F.}\ \bibnamefont
  {Kusmartseva}}, \bibinfo {author} {\bibfnamefont {A.}~\bibnamefont {Akrap}},
  \bibinfo {author} {\bibfnamefont {H.}~\bibnamefont {Berger}}, \bibinfo
  {author} {\bibfnamefont {L.}~\bibnamefont {Forr{\'{o}}}},\ and\ \bibinfo
  {author} {\bibfnamefont {E.}~\bibnamefont {Tuti{\v{s}}}},\ }\href
  {https://doi.org/10.1038/nmat2318} {\bibfield  {journal} {\bibinfo  {journal}
  {Nature Materials}\ }\textbf {\bibinfo {volume} {7}},\ \bibinfo {pages} {960}
  (\bibinfo {year} {2008})}\BibitemShut {NoStop}%
\bibitem [{\citenamefont {Yang}\ \emph {et~al.}(2012)\citenamefont {Yang},
  \citenamefont {Choi}, \citenamefont {Oh}, \citenamefont {Hogan},
  \citenamefont {Horibe}, \citenamefont {Kim}, \citenamefont {Min},\ and\
  \citenamefont {Cheong}}]{Yang2012}%
  \BibitemOpen
  \bibfield  {author} {\bibinfo {author} {\bibfnamefont {J.~J.}\ \bibnamefont
  {Yang}}, \bibinfo {author} {\bibfnamefont {Y.~J.}\ \bibnamefont {Choi}},
  \bibinfo {author} {\bibfnamefont {Y.~S.}\ \bibnamefont {Oh}}, \bibinfo
  {author} {\bibfnamefont {A.}~\bibnamefont {Hogan}}, \bibinfo {author}
  {\bibfnamefont {Y.}~\bibnamefont {Horibe}}, \bibinfo {author} {\bibfnamefont
  {K.}~\bibnamefont {Kim}}, \bibinfo {author} {\bibfnamefont {B.~I.}\
  \bibnamefont {Min}},\ and\ \bibinfo {author} {\bibfnamefont {S.-W.}\
  \bibnamefont {Cheong}},\ }\bibfield  {journal} {\bibinfo  {journal} {Physical
  Review Letters}\ }\textbf {\bibinfo {volume} {108}},\ \href
  {https://doi.org/10.1103/physrevlett.108.116402}
  {10.1103/physrevlett.108.116402} (\bibinfo {year} {2012})\BibitemShut
  {NoStop}%
\bibitem [{\citenamefont {Pyon}\ \emph {et~al.}(2012)\citenamefont {Pyon},
  \citenamefont {Kudo},\ and\ \citenamefont {Nohara}}]{Pyon2012}%
  \BibitemOpen
  \bibfield  {author} {\bibinfo {author} {\bibfnamefont {S.}~\bibnamefont
  {Pyon}}, \bibinfo {author} {\bibfnamefont {K.}~\bibnamefont {Kudo}},\ and\
  \bibinfo {author} {\bibfnamefont {M.}~\bibnamefont {Nohara}},\ }\href
  {https://doi.org/10.1143/jpsj.81.053701} {\bibfield  {journal} {\bibinfo
  {journal} {Journal of the Physical Society of Japan}\ }\textbf {\bibinfo
  {volume} {81}},\ \bibinfo {pages} {053701} (\bibinfo {year}
  {2012})}\BibitemShut {NoStop}%
\bibitem [{\citenamefont {Fang}\ \emph {et~al.}(2013)\citenamefont {Fang},
  \citenamefont {Xu}, \citenamefont {Dong}, \citenamefont {Zheng},\ and\
  \citenamefont {Wang}}]{Fang2013}%
  \BibitemOpen
  \bibfield  {author} {\bibinfo {author} {\bibfnamefont {A.~F.}\ \bibnamefont
  {Fang}}, \bibinfo {author} {\bibfnamefont {G.}~\bibnamefont {Xu}}, \bibinfo
  {author} {\bibfnamefont {T.}~\bibnamefont {Dong}}, \bibinfo {author}
  {\bibfnamefont {P.}~\bibnamefont {Zheng}},\ and\ \bibinfo {author}
  {\bibfnamefont {N.~L.}\ \bibnamefont {Wang}},\ }\bibfield  {journal}
  {\bibinfo  {journal} {Scientific Reports}\ }\textbf {\bibinfo {volume} {3}},\
  \href {https://doi.org/10.1038/srep01153} {10.1038/srep01153} (\bibinfo
  {year} {2013})\BibitemShut {NoStop}%
\bibitem [{\citenamefont {Kamitani}\ \emph {et~al.}(2016)\citenamefont
  {Kamitani}, \citenamefont {Sakai}, \citenamefont {Tokura},\ and\
  \citenamefont {Ishiwata}}]{Kamitani2016}%
  \BibitemOpen
  \bibfield  {author} {\bibinfo {author} {\bibfnamefont {M.}~\bibnamefont
  {Kamitani}}, \bibinfo {author} {\bibfnamefont {H.}~\bibnamefont {Sakai}},
  \bibinfo {author} {\bibfnamefont {Y.}~\bibnamefont {Tokura}},\ and\ \bibinfo
  {author} {\bibfnamefont {S.}~\bibnamefont {Ishiwata}},\ }\bibfield  {journal}
  {\bibinfo  {journal} {Physical Review B}\ }\textbf {\bibinfo {volume} {94}},\
  \href {https://doi.org/10.1103/physrevb.94.134507}
  {10.1103/physrevb.94.134507} (\bibinfo {year} {2016})\BibitemShut {NoStop}%
\bibitem [{\citenamefont {Kudo}\ \emph {et~al.}(2016)\citenamefont {Kudo},
  \citenamefont {Ishii},\ and\ \citenamefont {Nohara}}]{Kudo2016}%
  \BibitemOpen
  \bibfield  {author} {\bibinfo {author} {\bibfnamefont {K.}~\bibnamefont
  {Kudo}}, \bibinfo {author} {\bibfnamefont {H.}~\bibnamefont {Ishii}},\ and\
  \bibinfo {author} {\bibfnamefont {M.}~\bibnamefont {Nohara}},\ }\bibfield
  {journal} {\bibinfo  {journal} {Physical Review B}\ }\textbf {\bibinfo
  {volume} {93}},\ \href {https://doi.org/10.1103/physrevb.93.140505}
  {10.1103/physrevb.93.140505} (\bibinfo {year} {2016})\BibitemShut {NoStop}%
\bibitem [{\citenamefont {Heil}\ \emph {et~al.}(2017)\citenamefont {Heil},
  \citenamefont {Ponc{\'{e}}}, \citenamefont {Lambert}, \citenamefont
  {Schlipf}, \citenamefont {Margine},\ and\ \citenamefont
  {Giustino}}]{Heil2017}%
  \BibitemOpen
  \bibfield  {author} {\bibinfo {author} {\bibfnamefont {C.}~\bibnamefont
  {Heil}}, \bibinfo {author} {\bibfnamefont {S.}~\bibnamefont {Ponc{\'{e}}}},
  \bibinfo {author} {\bibfnamefont {H.}~\bibnamefont {Lambert}}, \bibinfo
  {author} {\bibfnamefont {M.}~\bibnamefont {Schlipf}}, \bibinfo {author}
  {\bibfnamefont {E.~R.}\ \bibnamefont {Margine}},\ and\ \bibinfo {author}
  {\bibfnamefont {F.}~\bibnamefont {Giustino}},\ }\bibfield  {journal}
  {\bibinfo  {journal} {Physical Review Letters}\ }\textbf {\bibinfo {volume}
  {119}},\ \href {https://doi.org/10.1103/physrevlett.119.087003}
  {10.1103/physrevlett.119.087003} (\bibinfo {year} {2017})\BibitemShut
  {NoStop}%
\bibitem [{\citenamefont {Saint-Paul}\ and\ \citenamefont
  {Monceau}(2020)}]{SaintPaul2020}%
  \BibitemOpen
  \bibfield  {author} {\bibinfo {author} {\bibfnamefont {M.}~\bibnamefont
  {Saint-Paul}}\ and\ \bibinfo {author} {\bibfnamefont {P.}~\bibnamefont
  {Monceau}},\ }\href {https://doi.org/10.1080/14786435.2020.1844917}
  {\bibfield  {journal} {\bibinfo  {journal} {Philosophical Magazine}\ ,\
  \bibinfo {pages} {1}} (\bibinfo {year} {2020})}\BibitemShut {NoStop}%
\bibitem [{\citenamefont {Chu}\ and\ \citenamefont {Testardi}(1974)}]{Chu1974}%
  \BibitemOpen
  \bibfield  {author} {\bibinfo {author} {\bibfnamefont {C.~W.}\ \bibnamefont
  {Chu}}\ and\ \bibinfo {author} {\bibfnamefont {L.~R.}\ \bibnamefont
  {Testardi}},\ }\href {https://doi.org/10.1103/physrevlett.32.766} {\bibfield
  {journal} {\bibinfo  {journal} {Physical Review Letters}\ }\textbf {\bibinfo
  {volume} {32}},\ \bibinfo {pages} {766} (\bibinfo {year} {1974})}\BibitemShut
  {NoStop}%
\bibitem [{\citenamefont {Chu}(1974)}]{Chu1974a}%
  \BibitemOpen
  \bibfield  {author} {\bibinfo {author} {\bibfnamefont {C.~W.}\ \bibnamefont
  {Chu}},\ }\href {https://doi.org/10.1103/physrevlett.33.1283} {\bibfield
  {journal} {\bibinfo  {journal} {Physical Review Letters}\ }\textbf {\bibinfo
  {volume} {33}},\ \bibinfo {pages} {1283} (\bibinfo {year}
  {1974})}\BibitemShut {NoStop}%
\bibitem [{\citenamefont {Testardi}(1975)}]{Testardi1975}%
  \BibitemOpen
  \bibfield  {author} {\bibinfo {author} {\bibfnamefont {L.~R.}\ \bibnamefont
  {Testardi}},\ }\href {https://doi.org/10.1103/revmodphys.47.637} {\bibfield
  {journal} {\bibinfo  {journal} {Reviews of Modern Physics}\ }\textbf
  {\bibinfo {volume} {47}},\ \bibinfo {pages} {637} (\bibinfo {year}
  {1975})}\BibitemShut {NoStop}%
\bibitem [{\citenamefont {Tanaka}\ \emph {et~al.}(2010)\citenamefont {Tanaka},
  \citenamefont {Miyake}, \citenamefont {Salce}, \citenamefont {Braithwaite},
  \citenamefont {Kagayama},\ and\ \citenamefont {Shimizu}}]{Tanaka2010}%
  \BibitemOpen
  \bibfield  {author} {\bibinfo {author} {\bibfnamefont {S.}~\bibnamefont
  {Tanaka}}, \bibinfo {author} {\bibfnamefont {A.}~\bibnamefont {Miyake}},
  \bibinfo {author} {\bibfnamefont {B.}~\bibnamefont {Salce}}, \bibinfo
  {author} {\bibfnamefont {D.}~\bibnamefont {Braithwaite}}, \bibinfo {author}
  {\bibfnamefont {T.}~\bibnamefont {Kagayama}},\ and\ \bibinfo {author}
  {\bibfnamefont {K.}~\bibnamefont {Shimizu}},\ }\href
  {https://doi.org/10.1088/1742-6596/200/1/012202} {\bibfield  {journal}
  {\bibinfo  {journal} {Journal of Physics: Conference Series}\ }\textbf
  {\bibinfo {volume} {200}},\ \bibinfo {pages} {012202} (\bibinfo {year}
  {2010})}\BibitemShut {NoStop}%
\bibitem [{\citenamefont {Bussmann-Holder}\ and\ \citenamefont
  {Bishop}(1992)}]{BussmannHolder1992}%
  \BibitemOpen
  \bibfield  {author} {\bibinfo {author} {\bibfnamefont {A.}~\bibnamefont
  {Bussmann-Holder}}\ and\ \bibinfo {author} {\bibfnamefont {A.~R.}\
  \bibnamefont {Bishop}},\ }\href {https://doi.org/10.1007/bf01313824}
  {\bibfield  {journal} {\bibinfo  {journal} {Zeitschrift f\"ur Physik B
  Condensed Matter}\ }\textbf {\bibinfo {volume} {86}},\ \bibinfo {pages} {183}
  (\bibinfo {year} {1992})}\BibitemShut {NoStop}%
\bibitem [{\citenamefont {Birgeneau}\ \emph {et~al.}(1987)\citenamefont
  {Birgeneau}, \citenamefont {Chen}, \citenamefont {Gabbe}, \citenamefont
  {Jenssen}, \citenamefont {Kastner}, \citenamefont {Peters}, \citenamefont
  {Picone}, \citenamefont {Thio}, \citenamefont {Thurston}, \citenamefont
  {Tuller}, \citenamefont {Axe}, \citenamefont {Böni},\ and\ \citenamefont
  {Shirane}}]{Birgeneau1987}%
  \BibitemOpen
  \bibfield  {author} {\bibinfo {author} {\bibfnamefont {R.~J.}\ \bibnamefont
  {Birgeneau}}, \bibinfo {author} {\bibfnamefont {C.~Y.}\ \bibnamefont {Chen}},
  \bibinfo {author} {\bibfnamefont {D.~R.}\ \bibnamefont {Gabbe}}, \bibinfo
  {author} {\bibfnamefont {H.~P.}\ \bibnamefont {Jenssen}}, \bibinfo {author}
  {\bibfnamefont {M.~A.}\ \bibnamefont {Kastner}}, \bibinfo {author}
  {\bibfnamefont {C.~J.}\ \bibnamefont {Peters}}, \bibinfo {author}
  {\bibfnamefont {P.~J.}\ \bibnamefont {Picone}}, \bibinfo {author}
  {\bibfnamefont {T.}~\bibnamefont {Thio}}, \bibinfo {author} {\bibfnamefont
  {T.~R.}\ \bibnamefont {Thurston}}, \bibinfo {author} {\bibfnamefont {H.~L.}\
  \bibnamefont {Tuller}}, \bibinfo {author} {\bibfnamefont {J.~D.}\
  \bibnamefont {Axe}}, \bibinfo {author} {\bibfnamefont {P.}~\bibnamefont
  {Böni}},\ and\ \bibinfo {author} {\bibfnamefont {G.}~\bibnamefont
  {Shirane}},\ }\href {https://doi.org/10.1103/physrevlett.59.1329} {\bibfield
  {journal} {\bibinfo  {journal} {Physical Review Letters}\ }\textbf {\bibinfo
  {volume} {59}},\ \bibinfo {pages} {1329} (\bibinfo {year}
  {1987})}\BibitemShut {NoStop}%
\bibitem [{\citenamefont {Wakimoto}\ \emph {et~al.}(2004)\citenamefont
  {Wakimoto}, \citenamefont {Lee}, \citenamefont {Gehring}, \citenamefont
  {Birgeneau},\ and\ \citenamefont {Shirane}}]{Wakimoto2004}%
  \BibitemOpen
  \bibfield  {author} {\bibinfo {author} {\bibfnamefont {S.}~\bibnamefont
  {Wakimoto}}, \bibinfo {author} {\bibfnamefont {S.}~\bibnamefont {Lee}},
  \bibinfo {author} {\bibfnamefont {P.~M.}\ \bibnamefont {Gehring}}, \bibinfo
  {author} {\bibfnamefont {R.~J.}\ \bibnamefont {Birgeneau}},\ and\ \bibinfo
  {author} {\bibfnamefont {G.}~\bibnamefont {Shirane}},\ }\href
  {https://doi.org/10.1143/jpsj.73.3413} {\bibfield  {journal} {\bibinfo
  {journal} {Journal of the Physical Society of Japan}\ }\textbf {\bibinfo
  {volume} {73}},\ \bibinfo {pages} {3413} (\bibinfo {year}
  {2004})}\BibitemShut {NoStop}%
\bibitem [{\citenamefont {de~la Cruz}\ \emph {et~al.}(2008)\citenamefont {de~la
  Cruz}, \citenamefont {Huang}, \citenamefont {Lynn}, \citenamefont {Li},
  \citenamefont {II}, \citenamefont {Zarestky}, \citenamefont {Mook},
  \citenamefont {Chen}, \citenamefont {Luo}, \citenamefont {Wang},\ and\
  \citenamefont {Dai}}]{Cruz2008}%
  \BibitemOpen
  \bibfield  {author} {\bibinfo {author} {\bibfnamefont {C.}~\bibnamefont
  {de~la Cruz}}, \bibinfo {author} {\bibfnamefont {Q.}~\bibnamefont {Huang}},
  \bibinfo {author} {\bibfnamefont {J.~W.}\ \bibnamefont {Lynn}}, \bibinfo
  {author} {\bibfnamefont {J.}~\bibnamefont {Li}}, \bibinfo {author}
  {\bibfnamefont {W.~R.}\ \bibnamefont {II}}, \bibinfo {author} {\bibfnamefont
  {J.~L.}\ \bibnamefont {Zarestky}}, \bibinfo {author} {\bibfnamefont {H.~A.}\
  \bibnamefont {Mook}}, \bibinfo {author} {\bibfnamefont {G.~F.}\ \bibnamefont
  {Chen}}, \bibinfo {author} {\bibfnamefont {J.~L.}\ \bibnamefont {Luo}},
  \bibinfo {author} {\bibfnamefont {N.~L.}\ \bibnamefont {Wang}},\ and\
  \bibinfo {author} {\bibfnamefont {P.}~\bibnamefont {Dai}},\ }\href
  {https://doi.org/10.1038/nature07057} {\bibfield  {journal} {\bibinfo
  {journal} {Nature}\ }\textbf {\bibinfo {volume} {453}},\ \bibinfo {pages}
  {899} (\bibinfo {year} {2008})}\BibitemShut {NoStop}%
\bibitem [{\citenamefont {Yoshizawa}\ \emph {et~al.}(2012)\citenamefont
  {Yoshizawa}, \citenamefont {Kimura}, \citenamefont {Chiba}, \citenamefont
  {Simayi}, \citenamefont {Nakanishi}, \citenamefont {Kihou}, \citenamefont
  {Lee}, \citenamefont {Iyo}, \citenamefont {Eisaki}, \citenamefont
  {Nakajima},\ and\ \citenamefont {ichi Uchida}}]{Yoshizawa2012}%
  \BibitemOpen
  \bibfield  {author} {\bibinfo {author} {\bibfnamefont {M.}~\bibnamefont
  {Yoshizawa}}, \bibinfo {author} {\bibfnamefont {D.}~\bibnamefont {Kimura}},
  \bibinfo {author} {\bibfnamefont {T.}~\bibnamefont {Chiba}}, \bibinfo
  {author} {\bibfnamefont {S.}~\bibnamefont {Simayi}}, \bibinfo {author}
  {\bibfnamefont {Y.}~\bibnamefont {Nakanishi}}, \bibinfo {author}
  {\bibfnamefont {K.}~\bibnamefont {Kihou}}, \bibinfo {author} {\bibfnamefont
  {C.-H.}\ \bibnamefont {Lee}}, \bibinfo {author} {\bibfnamefont
  {A.}~\bibnamefont {Iyo}}, \bibinfo {author} {\bibfnamefont {H.}~\bibnamefont
  {Eisaki}}, \bibinfo {author} {\bibfnamefont {M.}~\bibnamefont {Nakajima}},\
  and\ \bibinfo {author} {\bibfnamefont {S.}~\bibnamefont {ichi Uchida}},\
  }\href {https://doi.org/10.1143/jpsj.81.024604} {\bibfield  {journal}
  {\bibinfo  {journal} {Journal of the Physical Society of Japan}\ }\textbf
  {\bibinfo {volume} {81}},\ \bibinfo {pages} {024604} (\bibinfo {year}
  {2012})}\BibitemShut {NoStop}%
\bibitem [{\citenamefont {Niedziela}\ \emph {et~al.}(2011)\citenamefont
  {Niedziela}, \citenamefont {Parshall}, \citenamefont {Lokshin}, \citenamefont
  {Sefat}, \citenamefont {Alatas},\ and\ \citenamefont
  {Egami}}]{Niedziela2011}%
  \BibitemOpen
  \bibfield  {author} {\bibinfo {author} {\bibfnamefont {J.~L.}\ \bibnamefont
  {Niedziela}}, \bibinfo {author} {\bibfnamefont {D.}~\bibnamefont {Parshall}},
  \bibinfo {author} {\bibfnamefont {K.~A.}\ \bibnamefont {Lokshin}}, \bibinfo
  {author} {\bibfnamefont {A.~S.}\ \bibnamefont {Sefat}}, \bibinfo {author}
  {\bibfnamefont {A.}~\bibnamefont {Alatas}},\ and\ \bibinfo {author}
  {\bibfnamefont {T.}~\bibnamefont {Egami}},\ }\bibfield  {journal} {\bibinfo
  {journal} {Physical Review B}\ }\textbf {\bibinfo {volume} {84}},\ \href
  {https://doi.org/10.1103/physrevb.84.224305} {10.1103/physrevb.84.224305}
  (\bibinfo {year} {2011})\BibitemShut {NoStop}%
\bibitem [{\citenamefont {Kudo}\ \emph {et~al.}(2012)\citenamefont {Kudo},
  \citenamefont {Takasuga}, \citenamefont {Okamoto}, \citenamefont {Hiroi},\
  and\ \citenamefont {Nohara}}]{Kudo2012}%
  \BibitemOpen
  \bibfield  {author} {\bibinfo {author} {\bibfnamefont {K.}~\bibnamefont
  {Kudo}}, \bibinfo {author} {\bibfnamefont {M.}~\bibnamefont {Takasuga}},
  \bibinfo {author} {\bibfnamefont {Y.}~\bibnamefont {Okamoto}}, \bibinfo
  {author} {\bibfnamefont {Z.}~\bibnamefont {Hiroi}},\ and\ \bibinfo {author}
  {\bibfnamefont {M.}~\bibnamefont {Nohara}},\ }\bibfield  {journal} {\bibinfo
  {journal} {Physical Review Letters}\ }\textbf {\bibinfo {volume} {109}},\
  \href {https://doi.org/10.1103/physrevlett.109.097002}
  {10.1103/physrevlett.109.097002} (\bibinfo {year} {2012})\BibitemShut
  {NoStop}%
\bibitem [{\citenamefont {Hirai}\ \emph {et~al.}(2012)\citenamefont {Hirai},
  \citenamefont {von Rohr},\ and\ \citenamefont {Cava}}]{Hirai2012}%
  \BibitemOpen
  \bibfield  {author} {\bibinfo {author} {\bibfnamefont {D.}~\bibnamefont
  {Hirai}}, \bibinfo {author} {\bibfnamefont {F.}~\bibnamefont {von Rohr}},\
  and\ \bibinfo {author} {\bibfnamefont {R.~J.}\ \bibnamefont {Cava}},\
  }\bibfield  {journal} {\bibinfo  {journal} {Physical Review B}\ }\textbf
  {\bibinfo {volume} {86}},\ \href {https://doi.org/10.1103/physrevb.86.100505}
  {10.1103/physrevb.86.100505} (\bibinfo {year} {2012})\BibitemShut {NoStop}%
\bibitem [{\citenamefont {Kang}\ \emph {et~al.}(2011)\citenamefont {Kang},
  \citenamefont {Lee}, \citenamefont {Lynn}, \citenamefont {Shiryaev},\ and\
  \citenamefont {Barilo}}]{Kang2011}%
  \BibitemOpen
  \bibfield  {author} {\bibinfo {author} {\bibfnamefont {H.}~\bibnamefont
  {Kang}}, \bibinfo {author} {\bibfnamefont {Y.}~\bibnamefont {Lee}}, \bibinfo
  {author} {\bibfnamefont {J.}~\bibnamefont {Lynn}}, \bibinfo {author}
  {\bibfnamefont {S.}~\bibnamefont {Shiryaev}},\ and\ \bibinfo {author}
  {\bibfnamefont {S.}~\bibnamefont {Barilo}},\ }\href
  {https://doi.org/10.1016/j.physc.2011.02.010} {\bibfield  {journal} {\bibinfo
   {journal} {Physica C: Superconductivity}\ }\textbf {\bibinfo {volume}
  {471}},\ \bibinfo {pages} {303} (\bibinfo {year} {2011})}\BibitemShut
  {NoStop}%
\bibitem [{\citenamefont {Kase}\ \emph {et~al.}(2011)\citenamefont {Kase},
  \citenamefont {Hayamizu},\ and\ \citenamefont {Akimitsu}}]{Kase2011}%
  \BibitemOpen
  \bibfield  {author} {\bibinfo {author} {\bibfnamefont {N.}~\bibnamefont
  {Kase}}, \bibinfo {author} {\bibfnamefont {H.}~\bibnamefont {Hayamizu}},\
  and\ \bibinfo {author} {\bibfnamefont {J.}~\bibnamefont {Akimitsu}},\
  }\bibfield  {journal} {\bibinfo  {journal} {Physical Review B}\ }\textbf
  {\bibinfo {volume} {83}},\ \href {https://doi.org/10.1103/physrevb.83.184509}
  {10.1103/physrevb.83.184509} (\bibinfo {year} {2011})\BibitemShut {NoStop}%
\bibitem [{\citenamefont {Wang}\ and\ \citenamefont
  {Petrovic}(2012)}]{Wang2012}%
  \BibitemOpen
  \bibfield  {author} {\bibinfo {author} {\bibfnamefont {K.}~\bibnamefont
  {Wang}}\ and\ \bibinfo {author} {\bibfnamefont {C.}~\bibnamefont
  {Petrovic}},\ }\bibfield  {journal} {\bibinfo  {journal} {Physical Review B}\
  }\textbf {\bibinfo {volume} {86}},\ \href
  {https://doi.org/10.1103/physrevb.86.024522} {10.1103/physrevb.86.024522}
  (\bibinfo {year} {2012})\BibitemShut {NoStop}%
\bibitem [{\citenamefont {Klintberg}\ \emph {et~al.}(2012)\citenamefont
  {Klintberg}, \citenamefont {Goh}, \citenamefont {Alireza}, \citenamefont
  {Saines}, \citenamefont {Tompsett}, \citenamefont {Logg}, \citenamefont
  {Yang}, \citenamefont {Chen}, \citenamefont {Yoshimura},\ and\ \citenamefont
  {Grosche}}]{Klintberg2012}%
  \BibitemOpen
  \bibfield  {author} {\bibinfo {author} {\bibfnamefont {L.~E.}\ \bibnamefont
  {Klintberg}}, \bibinfo {author} {\bibfnamefont {S.~K.}\ \bibnamefont {Goh}},
  \bibinfo {author} {\bibfnamefont {P.~L.}\ \bibnamefont {Alireza}}, \bibinfo
  {author} {\bibfnamefont {P.~J.}\ \bibnamefont {Saines}}, \bibinfo {author}
  {\bibfnamefont {D.~A.}\ \bibnamefont {Tompsett}}, \bibinfo {author}
  {\bibfnamefont {P.~W.}\ \bibnamefont {Logg}}, \bibinfo {author}
  {\bibfnamefont {J.}~\bibnamefont {Yang}}, \bibinfo {author} {\bibfnamefont
  {B.}~\bibnamefont {Chen}}, \bibinfo {author} {\bibfnamefont {K.}~\bibnamefont
  {Yoshimura}},\ and\ \bibinfo {author} {\bibfnamefont {F.~M.}\ \bibnamefont
  {Grosche}},\ }\bibfield  {journal} {\bibinfo  {journal} {Physical Review
  Letters}\ }\textbf {\bibinfo {volume} {109}},\ \href
  {https://doi.org/10.1103/physrevlett.109.237008}
  {10.1103/physrevlett.109.237008} (\bibinfo {year} {2012})\BibitemShut
  {NoStop}%
\bibitem [{\citenamefont {Zhou}\ \emph {et~al.}(2012)\citenamefont {Zhou},
  \citenamefont {Zhang}, \citenamefont {Hong}, \citenamefont {Pan},
  \citenamefont {Qiu}, \citenamefont {Dong}, \citenamefont {Li},\ and\
  \citenamefont {Li}}]{Zhou2012}%
  \BibitemOpen
  \bibfield  {author} {\bibinfo {author} {\bibfnamefont {S.~Y.}\ \bibnamefont
  {Zhou}}, \bibinfo {author} {\bibfnamefont {H.}~\bibnamefont {Zhang}},
  \bibinfo {author} {\bibfnamefont {X.~C.}\ \bibnamefont {Hong}}, \bibinfo
  {author} {\bibfnamefont {B.~Y.}\ \bibnamefont {Pan}}, \bibinfo {author}
  {\bibfnamefont {X.}~\bibnamefont {Qiu}}, \bibinfo {author} {\bibfnamefont
  {W.~N.}\ \bibnamefont {Dong}}, \bibinfo {author} {\bibfnamefont {X.~L.}\
  \bibnamefont {Li}},\ and\ \bibinfo {author} {\bibfnamefont {S.~Y.}\
  \bibnamefont {Li}},\ }\bibfield  {journal} {\bibinfo  {journal} {Physical
  Review B}\ }\textbf {\bibinfo {volume} {86}},\ \href
  {https://doi.org/10.1103/physrevb.86.064504} {10.1103/physrevb.86.064504}
  (\bibinfo {year} {2012})\BibitemShut {NoStop}%
\bibitem [{\citenamefont {Biswas}\ \emph {et~al.}(2014)\citenamefont {Biswas},
  \citenamefont {Amato}, \citenamefont {Khasanov}, \citenamefont {Luetkens},
  \citenamefont {Wang}, \citenamefont {Petrovic}, \citenamefont {Cook},
  \citenamefont {Lees},\ and\ \citenamefont {Morenzoni}}]{Biswas2014}%
  \BibitemOpen
  \bibfield  {author} {\bibinfo {author} {\bibfnamefont {P.~K.}\ \bibnamefont
  {Biswas}}, \bibinfo {author} {\bibfnamefont {A.}~\bibnamefont {Amato}},
  \bibinfo {author} {\bibfnamefont {R.}~\bibnamefont {Khasanov}}, \bibinfo
  {author} {\bibfnamefont {H.}~\bibnamefont {Luetkens}}, \bibinfo {author}
  {\bibfnamefont {K.}~\bibnamefont {Wang}}, \bibinfo {author} {\bibfnamefont
  {C.}~\bibnamefont {Petrovic}}, \bibinfo {author} {\bibfnamefont {R.~M.}\
  \bibnamefont {Cook}}, \bibinfo {author} {\bibfnamefont {M.~R.}\ \bibnamefont
  {Lees}},\ and\ \bibinfo {author} {\bibfnamefont {E.}~\bibnamefont
  {Morenzoni}},\ }\bibfield  {journal} {\bibinfo  {journal} {Physical Review
  B}\ }\textbf {\bibinfo {volume} {90}},\ \href
  {https://doi.org/10.1103/physrevb.90.144505} {10.1103/physrevb.90.144505}
  (\bibinfo {year} {2014})\BibitemShut {NoStop}%
\bibitem [{\citenamefont {Goh}\ \emph {et~al.}(2015)\citenamefont {Goh},
  \citenamefont {Tompsett}, \citenamefont {Saines}, \citenamefont {Chang},
  \citenamefont {Matsumoto}, \citenamefont {Imai}, \citenamefont {Yoshimura},\
  and\ \citenamefont {Grosche}}]{Goh2015}%
  \BibitemOpen
  \bibfield  {author} {\bibinfo {author} {\bibfnamefont {S.}~\bibnamefont
  {Goh}}, \bibinfo {author} {\bibfnamefont {D.}~\bibnamefont {Tompsett}},
  \bibinfo {author} {\bibfnamefont {P.}~\bibnamefont {Saines}}, \bibinfo
  {author} {\bibfnamefont {H.}~\bibnamefont {Chang}}, \bibinfo {author}
  {\bibfnamefont {T.}~\bibnamefont {Matsumoto}}, \bibinfo {author}
  {\bibfnamefont {M.}~\bibnamefont {Imai}}, \bibinfo {author} {\bibfnamefont
  {K.}~\bibnamefont {Yoshimura}},\ and\ \bibinfo {author} {\bibfnamefont
  {F.}~\bibnamefont {Grosche}},\ }\bibfield  {journal} {\bibinfo  {journal}
  {Physical Review Letters}\ }\textbf {\bibinfo {volume} {114}},\ \href
  {https://doi.org/10.1103/physrevlett.114.097002}
  {10.1103/physrevlett.114.097002} (\bibinfo {year} {2015})\BibitemShut
  {NoStop}%
\bibitem [{\citenamefont {Gauzzi}\ \emph {et~al.}(2007)\citenamefont {Gauzzi},
  \citenamefont {Takashima}, \citenamefont {Takeshita}, \citenamefont
  {Terakura}, \citenamefont {Takagi}, \citenamefont {Emery}, \citenamefont
  {H{\'{e}}rold}, \citenamefont {Lagrange},\ and\ \citenamefont
  {Loupias}}]{Gauzzi2007}%
  \BibitemOpen
  \bibfield  {author} {\bibinfo {author} {\bibfnamefont {A.}~\bibnamefont
  {Gauzzi}}, \bibinfo {author} {\bibfnamefont {S.}~\bibnamefont {Takashima}},
  \bibinfo {author} {\bibfnamefont {N.}~\bibnamefont {Takeshita}}, \bibinfo
  {author} {\bibfnamefont {C.}~\bibnamefont {Terakura}}, \bibinfo {author}
  {\bibfnamefont {H.}~\bibnamefont {Takagi}}, \bibinfo {author} {\bibfnamefont
  {N.}~\bibnamefont {Emery}}, \bibinfo {author} {\bibfnamefont
  {C.}~\bibnamefont {H{\'{e}}rold}}, \bibinfo {author} {\bibfnamefont
  {P.}~\bibnamefont {Lagrange}},\ and\ \bibinfo {author} {\bibfnamefont
  {G.}~\bibnamefont {Loupias}},\ }\bibfield  {journal} {\bibinfo  {journal}
  {Physical Review Letters}\ }\textbf {\bibinfo {volume} {98}},\ \href
  {https://doi.org/10.1103/physrevlett.98.067002}
  {10.1103/physrevlett.98.067002} (\bibinfo {year} {2007})\BibitemShut
  {NoStop}%
\bibitem [{\citenamefont {Lubczynski}\ \emph {et~al.}(1996)\citenamefont
  {Lubczynski}, \citenamefont {Demishev}, \citenamefont {Singleton},
  \citenamefont {Caulfield}, \citenamefont {du~Croo~de Jongh}, \citenamefont
  {Kepert}, \citenamefont {Blundell}, \citenamefont {Hayes}, \citenamefont
  {Kurmoo},\ and\ \citenamefont {Day}}]{Lubczynski1996}%
  \BibitemOpen
  \bibfield  {author} {\bibinfo {author} {\bibfnamefont {W.}~\bibnamefont
  {Lubczynski}}, \bibinfo {author} {\bibfnamefont {S.~V.}\ \bibnamefont
  {Demishev}}, \bibinfo {author} {\bibfnamefont {J.}~\bibnamefont {Singleton}},
  \bibinfo {author} {\bibfnamefont {J.~M.}\ \bibnamefont {Caulfield}}, \bibinfo
  {author} {\bibfnamefont {L.}~\bibnamefont {du~Croo~de Jongh}}, \bibinfo
  {author} {\bibfnamefont {C.~J.}\ \bibnamefont {Kepert}}, \bibinfo {author}
  {\bibfnamefont {S.~J.}\ \bibnamefont {Blundell}}, \bibinfo {author}
  {\bibfnamefont {W.}~\bibnamefont {Hayes}}, \bibinfo {author} {\bibfnamefont
  {M.}~\bibnamefont {Kurmoo}},\ and\ \bibinfo {author} {\bibfnamefont
  {P.}~\bibnamefont {Day}},\ }\href
  {https://doi.org/10.1088/0953-8984/8/33/009} {\bibfield  {journal} {\bibinfo
  {journal} {Journal of Physics: Condensed Matter}\ }\textbf {\bibinfo {volume}
  {8}},\ \bibinfo {pages} {6005} (\bibinfo {year} {1996})}\BibitemShut
  {NoStop}%
\bibitem [{\citenamefont {Wosnitza}(2001)}]{Wosnitza2001}%
  \BibitemOpen
  \bibfield  {author} {\bibinfo {author} {\bibfnamefont {J.}~\bibnamefont
  {Wosnitza}},\ }\href {https://doi.org/10.1016/s1359-0286(00)00039-5}
  {\bibfield  {journal} {\bibinfo  {journal} {Current Opinion in Solid State
  and Materials Science}\ }\textbf {\bibinfo {volume} {5}},\ \bibinfo {pages}
  {131} (\bibinfo {year} {2001})}\BibitemShut {NoStop}%
\bibitem [{\citenamefont {Degtyareva}\ \emph {et~al.}(2007)\citenamefont
  {Degtyareva}, \citenamefont {Magnitskaya}, \citenamefont {Kohanoff},
  \citenamefont {Profeta}, \citenamefont {Scandolo}, \citenamefont {Hanfland},
  \citenamefont {McMahon},\ and\ \citenamefont {Gregoryanz}}]{Degtyareva2007}%
  \BibitemOpen
  \bibfield  {author} {\bibinfo {author} {\bibfnamefont {O.}~\bibnamefont
  {Degtyareva}}, \bibinfo {author} {\bibfnamefont {M.~V.}\ \bibnamefont
  {Magnitskaya}}, \bibinfo {author} {\bibfnamefont {J.}~\bibnamefont
  {Kohanoff}}, \bibinfo {author} {\bibfnamefont {G.}~\bibnamefont {Profeta}},
  \bibinfo {author} {\bibfnamefont {S.}~\bibnamefont {Scandolo}}, \bibinfo
  {author} {\bibfnamefont {M.}~\bibnamefont {Hanfland}}, \bibinfo {author}
  {\bibfnamefont {M.~I.}\ \bibnamefont {McMahon}},\ and\ \bibinfo {author}
  {\bibfnamefont {E.}~\bibnamefont {Gregoryanz}},\ }\bibfield  {journal}
  {\bibinfo  {journal} {Physical Review Letters}\ }\textbf {\bibinfo {volume}
  {99}},\ \href {https://doi.org/10.1103/physrevlett.99.155505}
  {10.1103/physrevlett.99.155505} (\bibinfo {year} {2007})\BibitemShut
  {NoStop}%
\bibitem [{\citenamefont {Yu}\ \emph {et~al.}(2015)\citenamefont {Yu},
  \citenamefont {Cheung}, \citenamefont {Saines}, \citenamefont {Imai},
  \citenamefont {Matsumoto}, \citenamefont {Michioka}, \citenamefont
  {Yoshimura},\ and\ \citenamefont {Goh}}]{Yu2015}%
  \BibitemOpen
  \bibfield  {author} {\bibinfo {author} {\bibfnamefont {W.~C.}\ \bibnamefont
  {Yu}}, \bibinfo {author} {\bibfnamefont {Y.~W.}\ \bibnamefont {Cheung}},
  \bibinfo {author} {\bibfnamefont {P.~J.}\ \bibnamefont {Saines}}, \bibinfo
  {author} {\bibfnamefont {M.}~\bibnamefont {Imai}}, \bibinfo {author}
  {\bibfnamefont {T.}~\bibnamefont {Matsumoto}}, \bibinfo {author}
  {\bibfnamefont {C.}~\bibnamefont {Michioka}}, \bibinfo {author}
  {\bibfnamefont {K.}~\bibnamefont {Yoshimura}},\ and\ \bibinfo {author}
  {\bibfnamefont {S.~K.}\ \bibnamefont {Goh}},\ }\bibfield  {journal} {\bibinfo
   {journal} {Physical Review Letters}\ }\textbf {\bibinfo {volume} {115}},\
  \href {https://doi.org/10.1103/physrevlett.115.207003}
  {10.1103/physrevlett.115.207003} (\bibinfo {year} {2015})\BibitemShut
  {NoStop}%
\bibitem [{\citenamefont {Cheung}\ \emph {et~al.}(2018)\citenamefont {Cheung},
  \citenamefont {Hu}, \citenamefont {Imai}, \citenamefont {Tanioku},
  \citenamefont {Kanagawa}, \citenamefont {Murakawa}, \citenamefont {Moriyama},
  \citenamefont {Zhang}, \citenamefont {Lai}, \citenamefont {Yoshimura},
  \citenamefont {Grosche}, \citenamefont {Kaneko}, \citenamefont {Tsutsui},\
  and\ \citenamefont {Goh}}]{Cheung2018}%
  \BibitemOpen
  \bibfield  {author} {\bibinfo {author} {\bibfnamefont {Y.~W.}\ \bibnamefont
  {Cheung}}, \bibinfo {author} {\bibfnamefont {Y.~J.}\ \bibnamefont {Hu}},
  \bibinfo {author} {\bibfnamefont {M.}~\bibnamefont {Imai}}, \bibinfo {author}
  {\bibfnamefont {Y.}~\bibnamefont {Tanioku}}, \bibinfo {author} {\bibfnamefont
  {H.}~\bibnamefont {Kanagawa}}, \bibinfo {author} {\bibfnamefont
  {J.}~\bibnamefont {Murakawa}}, \bibinfo {author} {\bibfnamefont
  {K.}~\bibnamefont {Moriyama}}, \bibinfo {author} {\bibfnamefont
  {W.}~\bibnamefont {Zhang}}, \bibinfo {author} {\bibfnamefont {K.~T.}\
  \bibnamefont {Lai}}, \bibinfo {author} {\bibfnamefont {K.}~\bibnamefont
  {Yoshimura}}, \bibinfo {author} {\bibfnamefont {F.~M.}\ \bibnamefont
  {Grosche}}, \bibinfo {author} {\bibfnamefont {K.}~\bibnamefont {Kaneko}},
  \bibinfo {author} {\bibfnamefont {S.}~\bibnamefont {Tsutsui}},\ and\ \bibinfo
  {author} {\bibfnamefont {S.~K.}\ \bibnamefont {Goh}},\ }\bibfield  {journal}
  {\bibinfo  {journal} {Physical Review B}\ }\textbf {\bibinfo {volume} {98}},\
  \href {https://doi.org/10.1103/physrevb.98.161103}
  {10.1103/physrevb.98.161103} (\bibinfo {year} {2018})\BibitemShut {NoStop}%
\bibitem [{\citenamefont {Coleman}\ \emph {et~al.}(1973)\citenamefont
  {Coleman}, \citenamefont {Cohen}, \citenamefont {Sandman}, \citenamefont
  {Yamagishi}, \citenamefont {Garito},\ and\ \citenamefont
  {Heeger}}]{Coleman1973}%
  \BibitemOpen
  \bibfield  {author} {\bibinfo {author} {\bibfnamefont {L.}~\bibnamefont
  {Coleman}}, \bibinfo {author} {\bibfnamefont {M.}~\bibnamefont {Cohen}},
  \bibinfo {author} {\bibfnamefont {D.}~\bibnamefont {Sandman}}, \bibinfo
  {author} {\bibfnamefont {F.}~\bibnamefont {Yamagishi}}, \bibinfo {author}
  {\bibfnamefont {A.}~\bibnamefont {Garito}},\ and\ \bibinfo {author}
  {\bibfnamefont {A.}~\bibnamefont {Heeger}},\ }\href
  {https://doi.org/10.1016/0038-1098(73)90127-0} {\bibfield  {journal}
  {\bibinfo  {journal} {Solid State Communications}\ }\textbf {\bibinfo
  {volume} {12}},\ \bibinfo {pages} {1125} (\bibinfo {year}
  {1973})}\BibitemShut {NoStop}%
\bibitem [{\citenamefont {Edwards}\ \emph {et~al.}(1994)\citenamefont
  {Edwards}, \citenamefont {Barr}, \citenamefont {Markert},\ and\ \citenamefont
  {de~Lozanne}}]{Edwards1994}%
  \BibitemOpen
  \bibfield  {author} {\bibinfo {author} {\bibfnamefont {H.~L.}\ \bibnamefont
  {Edwards}}, \bibinfo {author} {\bibfnamefont {A.~L.}\ \bibnamefont {Barr}},
  \bibinfo {author} {\bibfnamefont {J.~T.}\ \bibnamefont {Markert}},\ and\
  \bibinfo {author} {\bibfnamefont {A.~L.}\ \bibnamefont {de~Lozanne}},\ }\href
  {https://doi.org/10.1103/physrevlett.73.1154} {\bibfield  {journal} {\bibinfo
   {journal} {Physical Review Letters}\ }\textbf {\bibinfo {volume} {73}},\
  \bibinfo {pages} {1154} (\bibinfo {year} {1994})}\BibitemShut {NoStop}%
\bibitem [{\citenamefont {Gabovich}\ \emph {et~al.}(2001)\citenamefont
  {Gabovich}, \citenamefont {Voitenko}, \citenamefont {Annett},\ and\
  \citenamefont {Ausloos}}]{Gabovich2001}%
  \BibitemOpen
  \bibfield  {author} {\bibinfo {author} {\bibfnamefont {A.~M.}\ \bibnamefont
  {Gabovich}}, \bibinfo {author} {\bibfnamefont {A.~I.}\ \bibnamefont
  {Voitenko}}, \bibinfo {author} {\bibfnamefont {J.~F.}\ \bibnamefont
  {Annett}},\ and\ \bibinfo {author} {\bibfnamefont {M.}~\bibnamefont
  {Ausloos}},\ }\href {https://doi.org/10.1088/0953-2048/14/4/201} {\bibfield
  {journal} {\bibinfo  {journal} {Superconductor Science and Technology}\
  }\textbf {\bibinfo {volume} {14}},\ \bibinfo {pages} {R1} (\bibinfo {year}
  {2001})}\BibitemShut {NoStop}%
\bibitem [{\citenamefont {Gabovich}\ \emph {et~al.}(2002)\citenamefont
  {Gabovich}, \citenamefont {Voitenko},\ and\ \citenamefont
  {Ausloos}}]{Gabovich2002}%
  \BibitemOpen
  \bibfield  {author} {\bibinfo {author} {\bibfnamefont {A.}~\bibnamefont
  {Gabovich}}, \bibinfo {author} {\bibfnamefont {A.}~\bibnamefont {Voitenko}},\
  and\ \bibinfo {author} {\bibfnamefont {M.}~\bibnamefont {Ausloos}},\ }\href
  {https://doi.org/10.1016/s0370-1573(02)00029-7} {\bibfield  {journal}
  {\bibinfo  {journal} {Physics Reports}\ }\textbf {\bibinfo {volume} {367}},\
  \bibinfo {pages} {583} (\bibinfo {year} {2002})}\BibitemShut {NoStop}%
\bibitem [{\citenamefont {Dagotto}(2005)}]{Dagotto2005}%
  \BibitemOpen
  \bibfield  {author} {\bibinfo {author} {\bibfnamefont {E.}~\bibnamefont
  {Dagotto}},\ }\href {https://doi.org/10.1126/science.1107559} {\bibfield
  {journal} {\bibinfo  {journal} {Science}\ }\textbf {\bibinfo {volume}
  {309}},\ \bibinfo {pages} {257} (\bibinfo {year} {2005})}\BibitemShut
  {NoStop}%
\bibitem [{\citenamefont {Stewart}(2017)}]{Stewart2017}%
  \BibitemOpen
  \bibfield  {author} {\bibinfo {author} {\bibfnamefont {G.~R.}\ \bibnamefont
  {Stewart}},\ }\href {https://doi.org/10.1080/00018732.2017.1331615}
  {\bibfield  {journal} {\bibinfo  {journal} {Advances in Physics}\ }\textbf
  {\bibinfo {volume} {66}},\ \bibinfo {pages} {75} (\bibinfo {year}
  {2017})}\BibitemShut {NoStop}%
\bibitem [{\citenamefont {Mathur}\ \emph {et~al.}(1998)\citenamefont {Mathur},
  \citenamefont {Grosche}, \citenamefont {Julian}, \citenamefont {Walker},
  \citenamefont {Freye}, \citenamefont {Haselwimmer},\ and\ \citenamefont
  {Lonzarich}}]{Mathur1998}%
  \BibitemOpen
  \bibfield  {author} {\bibinfo {author} {\bibfnamefont {N.~D.}\ \bibnamefont
  {Mathur}}, \bibinfo {author} {\bibfnamefont {F.~M.}\ \bibnamefont {Grosche}},
  \bibinfo {author} {\bibfnamefont {S.~R.}\ \bibnamefont {Julian}}, \bibinfo
  {author} {\bibfnamefont {I.~R.}\ \bibnamefont {Walker}}, \bibinfo {author}
  {\bibfnamefont {D.~M.}\ \bibnamefont {Freye}}, \bibinfo {author}
  {\bibfnamefont {R.~K.~W.}\ \bibnamefont {Haselwimmer}},\ and\ \bibinfo
  {author} {\bibfnamefont {G.~G.}\ \bibnamefont {Lonzarich}},\ }\href
  {https://doi.org/10.1038/27838} {\bibfield  {journal} {\bibinfo  {journal}
  {Nature}\ }\textbf {\bibinfo {volume} {394}},\ \bibinfo {pages} {39}
  (\bibinfo {year} {1998})}\BibitemShut {NoStop}%
\bibitem [{\citenamefont {Gegenwart}\ \emph {et~al.}(2008)\citenamefont
  {Gegenwart}, \citenamefont {Si},\ and\ \citenamefont
  {Steglich}}]{Gegenwart2008}%
  \BibitemOpen
  \bibfield  {author} {\bibinfo {author} {\bibfnamefont {P.}~\bibnamefont
  {Gegenwart}}, \bibinfo {author} {\bibfnamefont {Q.}~\bibnamefont {Si}},\ and\
  \bibinfo {author} {\bibfnamefont {F.}~\bibnamefont {Steglich}},\ }\href
  {https://doi.org/10.1038/nphys892} {\bibfield  {journal} {\bibinfo  {journal}
  {Nature Physics}\ }\textbf {\bibinfo {volume} {4}},\ \bibinfo {pages} {186}
  (\bibinfo {year} {2008})}\BibitemShut {NoStop}%
\bibitem [{\citenamefont {Paglione}\ and\ \citenamefont
  {Greene}(2010)}]{Paglione2010}%
  \BibitemOpen
  \bibfield  {author} {\bibinfo {author} {\bibfnamefont {J.}~\bibnamefont
  {Paglione}}\ and\ \bibinfo {author} {\bibfnamefont {R.~L.}\ \bibnamefont
  {Greene}},\ }\href {https://doi.org/10.1038/nphys1759} {\bibfield  {journal}
  {\bibinfo  {journal} {Nature Physics}\ }\textbf {\bibinfo {volume} {6}},\
  \bibinfo {pages} {645} (\bibinfo {year} {2010})}\BibitemShut {NoStop}%
\bibitem [{\citenamefont {Ishida}\ \emph {et~al.}(2009)\citenamefont {Ishida},
  \citenamefont {Nakai},\ and\ \citenamefont {Hosono}}]{Ishida2009}%
  \BibitemOpen
  \bibfield  {author} {\bibinfo {author} {\bibfnamefont {K.}~\bibnamefont
  {Ishida}}, \bibinfo {author} {\bibfnamefont {Y.}~\bibnamefont {Nakai}},\ and\
  \bibinfo {author} {\bibfnamefont {H.}~\bibnamefont {Hosono}},\ }\href
  {https://doi.org/10.1143/jpsj.78.062001} {\bibfield  {journal} {\bibinfo
  {journal} {Journal of the Physical Society of Japan}\ }\textbf {\bibinfo
  {volume} {78}},\ \bibinfo {pages} {062001} (\bibinfo {year}
  {2009})}\BibitemShut {NoStop}%
\bibitem [{\citenamefont {Hashimoto}\ \emph {et~al.}(2012)\citenamefont
  {Hashimoto}, \citenamefont {Cho}, \citenamefont {Shibauchi}, \citenamefont
  {Kasahara}, \citenamefont {Mizukami}, \citenamefont {Katsumata},
  \citenamefont {Tsuruhara}, \citenamefont {Terashima}, \citenamefont {Ikeda},
  \citenamefont {Tanatar}, \citenamefont {Kitano}, \citenamefont {Salovich},
  \citenamefont {Giannetta}, \citenamefont {Walmsley}, \citenamefont
  {Carrington}, \citenamefont {Prozorov},\ and\ \citenamefont
  {Matsuda}}]{Hashimoto2012}%
  \BibitemOpen
  \bibfield  {author} {\bibinfo {author} {\bibfnamefont {K.}~\bibnamefont
  {Hashimoto}}, \bibinfo {author} {\bibfnamefont {K.}~\bibnamefont {Cho}},
  \bibinfo {author} {\bibfnamefont {T.}~\bibnamefont {Shibauchi}}, \bibinfo
  {author} {\bibfnamefont {S.}~\bibnamefont {Kasahara}}, \bibinfo {author}
  {\bibfnamefont {Y.}~\bibnamefont {Mizukami}}, \bibinfo {author}
  {\bibfnamefont {R.}~\bibnamefont {Katsumata}}, \bibinfo {author}
  {\bibfnamefont {Y.}~\bibnamefont {Tsuruhara}}, \bibinfo {author}
  {\bibfnamefont {T.}~\bibnamefont {Terashima}}, \bibinfo {author}
  {\bibfnamefont {H.}~\bibnamefont {Ikeda}}, \bibinfo {author} {\bibfnamefont
  {M.~A.}\ \bibnamefont {Tanatar}}, \bibinfo {author} {\bibfnamefont
  {H.}~\bibnamefont {Kitano}}, \bibinfo {author} {\bibfnamefont
  {N.}~\bibnamefont {Salovich}}, \bibinfo {author} {\bibfnamefont {R.~W.}\
  \bibnamefont {Giannetta}}, \bibinfo {author} {\bibfnamefont {P.}~\bibnamefont
  {Walmsley}}, \bibinfo {author} {\bibfnamefont {A.}~\bibnamefont
  {Carrington}}, \bibinfo {author} {\bibfnamefont {R.}~\bibnamefont
  {Prozorov}},\ and\ \bibinfo {author} {\bibfnamefont {Y.}~\bibnamefont
  {Matsuda}},\ }\href {https://doi.org/10.1126/science.1219821} {\bibfield
  {journal} {\bibinfo  {journal} {Science}\ }\textbf {\bibinfo {volume}
  {336}},\ \bibinfo {pages} {1554} (\bibinfo {year} {2012})}\BibitemShut
  {NoStop}%
\bibitem [{\citenamefont {Shibauchi}\ \emph {et~al.}(2014)\citenamefont
  {Shibauchi}, \citenamefont {Carrington},\ and\ \citenamefont
  {Matsuda}}]{Shibauchi2014}%
  \BibitemOpen
  \bibfield  {author} {\bibinfo {author} {\bibfnamefont {T.}~\bibnamefont
  {Shibauchi}}, \bibinfo {author} {\bibfnamefont {A.}~\bibnamefont
  {Carrington}},\ and\ \bibinfo {author} {\bibfnamefont {Y.}~\bibnamefont
  {Matsuda}},\ }\href
  {https://doi.org/10.1146/annurev-conmatphys-031113-133921} {\bibfield
  {journal} {\bibinfo  {journal} {Annual Review of Condensed Matter Physics}\
  }\textbf {\bibinfo {volume} {5}},\ \bibinfo {pages} {113} (\bibinfo {year}
  {2014})}\BibitemShut {NoStop}%
\bibitem [{\citenamefont {Bardeen}\ \emph
  {et~al.}(1957{\natexlab{a}})\citenamefont {Bardeen}, \citenamefont {Cooper},\
  and\ \citenamefont {Schrieffer}}]{Bardeen1957}%
  \BibitemOpen
  \bibfield  {author} {\bibinfo {author} {\bibfnamefont {J.}~\bibnamefont
  {Bardeen}}, \bibinfo {author} {\bibfnamefont {L.~N.}\ \bibnamefont
  {Cooper}},\ and\ \bibinfo {author} {\bibfnamefont {J.~R.}\ \bibnamefont
  {Schrieffer}},\ }\href {https://doi.org/10.1103/physrev.106.162} {\bibfield
  {journal} {\bibinfo  {journal} {Physical Review}\ }\textbf {\bibinfo {volume}
  {106}},\ \bibinfo {pages} {162} (\bibinfo {year}
  {1957}{\natexlab{a}})}\BibitemShut {NoStop}%
\bibitem [{\citenamefont {Bardeen}\ \emph
  {et~al.}(1957{\natexlab{b}})\citenamefont {Bardeen}, \citenamefont {Cooper},\
  and\ \citenamefont {Schrieffer}}]{Bardeen1957a}%
  \BibitemOpen
  \bibfield  {author} {\bibinfo {author} {\bibfnamefont {J.}~\bibnamefont
  {Bardeen}}, \bibinfo {author} {\bibfnamefont {L.~N.}\ \bibnamefont
  {Cooper}},\ and\ \bibinfo {author} {\bibfnamefont {J.~R.}\ \bibnamefont
  {Schrieffer}},\ }\href {https://doi.org/10.1103/physrev.108.1175} {\bibfield
  {journal} {\bibinfo  {journal} {Physical Review}\ }\textbf {\bibinfo {volume}
  {108}},\ \bibinfo {pages} {1175} (\bibinfo {year}
  {1957}{\natexlab{b}})}\BibitemShut {NoStop}%
\bibitem [{\citenamefont {Hayamizu}\ \emph {et~al.}(2011)\citenamefont
  {Hayamizu}, \citenamefont {Kase},\ and\ \citenamefont
  {Akimitsu}}]{Hayamizu2011}%
  \BibitemOpen
  \bibfield  {author} {\bibinfo {author} {\bibfnamefont {H.}~\bibnamefont
  {Hayamizu}}, \bibinfo {author} {\bibfnamefont {N.}~\bibnamefont {Kase}},\
  and\ \bibinfo {author} {\bibfnamefont {J.}~\bibnamefont {Akimitsu}},\ }\href
  {https://doi.org/10.1143/jpsjs.80sa.sa114} {\bibfield  {journal} {\bibinfo
  {journal} {Journal of the Physical Society of Japan}\ }\textbf {\bibinfo
  {volume} {80}},\ \bibinfo {pages} {SA114} (\bibinfo {year}
  {2011})}\BibitemShut {NoStop}%
\bibitem [{\citenamefont {Shen}\ \emph {et~al.}(2020)\citenamefont {Shen},
  \citenamefont {Du}, \citenamefont {Li}, \citenamefont {Thamizhavel},
  \citenamefont {Smidman}, \citenamefont {Nie}, \citenamefont {Luo},
  \citenamefont {Le}, \citenamefont {Hossain},\ and\ \citenamefont
  {Yuan}}]{Shen2020}%
  \BibitemOpen
  \bibfield  {author} {\bibinfo {author} {\bibfnamefont {B.}~\bibnamefont
  {Shen}}, \bibinfo {author} {\bibfnamefont {F.}~\bibnamefont {Du}}, \bibinfo
  {author} {\bibfnamefont {R.}~\bibnamefont {Li}}, \bibinfo {author}
  {\bibfnamefont {A.}~\bibnamefont {Thamizhavel}}, \bibinfo {author}
  {\bibfnamefont {M.}~\bibnamefont {Smidman}}, \bibinfo {author} {\bibfnamefont
  {Z.~Y.}\ \bibnamefont {Nie}}, \bibinfo {author} {\bibfnamefont {S.~S.}\
  \bibnamefont {Luo}}, \bibinfo {author} {\bibfnamefont {T.}~\bibnamefont
  {Le}}, \bibinfo {author} {\bibfnamefont {Z.}~\bibnamefont {Hossain}},\ and\
  \bibinfo {author} {\bibfnamefont {H.~Q.}\ \bibnamefont {Yuan}},\ }\bibfield
  {journal} {\bibinfo  {journal} {Physical Review B}\ }\textbf {\bibinfo
  {volume} {101}},\ \href {https://doi.org/10.1103/physrevb.101.144501}
  {10.1103/physrevb.101.144501} (\bibinfo {year} {2020})\BibitemShut {NoStop}%
\bibitem [{\citenamefont {Poudel}\ \emph {et~al.}(2016)\citenamefont {Poudel},
  \citenamefont {May}, \citenamefont {Koehler}, \citenamefont {McGuire},
  \citenamefont {Mukhopadhyay}, \citenamefont {Calder}, \citenamefont
  {Baumbach}, \citenamefont {Mukherjee}, \citenamefont {Sapkota}, \citenamefont
  {de~la Cruz}, \citenamefont {Singh}, \citenamefont {Mandrus},\ and\
  \citenamefont {Christianson}}]{Poudel2016}%
  \BibitemOpen
  \bibfield  {author} {\bibinfo {author} {\bibfnamefont {L.}~\bibnamefont
  {Poudel}}, \bibinfo {author} {\bibfnamefont {A.}~\bibnamefont {May}},
  \bibinfo {author} {\bibfnamefont {M.}~\bibnamefont {Koehler}}, \bibinfo
  {author} {\bibfnamefont {M.}~\bibnamefont {McGuire}}, \bibinfo {author}
  {\bibfnamefont {S.}~\bibnamefont {Mukhopadhyay}}, \bibinfo {author}
  {\bibfnamefont {S.}~\bibnamefont {Calder}}, \bibinfo {author} {\bibfnamefont
  {R.}~\bibnamefont {Baumbach}}, \bibinfo {author} {\bibfnamefont
  {R.}~\bibnamefont {Mukherjee}}, \bibinfo {author} {\bibfnamefont
  {D.}~\bibnamefont {Sapkota}}, \bibinfo {author} {\bibfnamefont
  {C.}~\bibnamefont {de~la Cruz}}, \bibinfo {author} {\bibfnamefont
  {D.}~\bibnamefont {Singh}}, \bibinfo {author} {\bibfnamefont
  {D.}~\bibnamefont {Mandrus}},\ and\ \bibinfo {author} {\bibfnamefont
  {A.}~\bibnamefont {Christianson}},\ }\bibfield  {journal} {\bibinfo
  {journal} {Physical Review Letters}\ }\textbf {\bibinfo {volume} {117}},\
  \href {https://doi.org/10.1103/physrevlett.117.235701}
  {10.1103/physrevlett.117.235701} (\bibinfo {year} {2016})\BibitemShut
  {NoStop}%
\bibitem [{\citenamefont {Carneiro}\ \emph {et~al.}(2020)\citenamefont
  {Carneiro}, \citenamefont {Veiga}, \citenamefont {Mardegan}, \citenamefont
  {Khan}, \citenamefont {Macchiutti}, \citenamefont {L{\'{o}}pez},\ and\
  \citenamefont {Bittar}}]{Carneiro2020}%
  \BibitemOpen
  \bibfield  {author} {\bibinfo {author} {\bibfnamefont {F.~B.}\ \bibnamefont
  {Carneiro}}, \bibinfo {author} {\bibfnamefont {L.~S.~I.}\ \bibnamefont
  {Veiga}}, \bibinfo {author} {\bibfnamefont {J.~R.~L.}\ \bibnamefont
  {Mardegan}}, \bibinfo {author} {\bibfnamefont {R.}~\bibnamefont {Khan}},
  \bibinfo {author} {\bibfnamefont {C.}~\bibnamefont {Macchiutti}}, \bibinfo
  {author} {\bibfnamefont {A.}~\bibnamefont {L{\'{o}}pez}},\ and\ \bibinfo
  {author} {\bibfnamefont {E.~M.}\ \bibnamefont {Bittar}},\ }\bibfield
  {journal} {\bibinfo  {journal} {Physical Review B}\ }\textbf {\bibinfo
  {volume} {101}},\ \href {https://doi.org/10.1103/physrevb.101.195135}
  {10.1103/physrevb.101.195135} (\bibinfo {year} {2020})\BibitemShut {NoStop}%
\bibitem [{\citenamefont {Kim}\ \emph {et~al.}(2015)\citenamefont {Kim},
  \citenamefont {Kim},\ and\ \citenamefont {Min}}]{Kim2015}%
  \BibitemOpen
  \bibfield  {author} {\bibinfo {author} {\bibfnamefont {S.}~\bibnamefont
  {Kim}}, \bibinfo {author} {\bibfnamefont {K.}~\bibnamefont {Kim}},\ and\
  \bibinfo {author} {\bibfnamefont {B.~I.}\ \bibnamefont {Min}},\ }\bibfield
  {journal} {\bibinfo  {journal} {Scientific Reports}\ }\textbf {\bibinfo
  {volume} {5}},\ \href {https://doi.org/10.1038/srep15052} {10.1038/srep15052}
  (\bibinfo {year} {2015})\BibitemShut {NoStop}%
\bibitem [{\citenamefont {Biswas}\ \emph {et~al.}(2015)\citenamefont {Biswas},
  \citenamefont {Guguchia}, \citenamefont {Khasanov}, \citenamefont {Chinotti},
  \citenamefont {Li}, \citenamefont {Wang}, \citenamefont {Petrovic},\ and\
  \citenamefont {Morenzoni}}]{Biswas2015}%
  \BibitemOpen
  \bibfield  {author} {\bibinfo {author} {\bibfnamefont {P.~K.}\ \bibnamefont
  {Biswas}}, \bibinfo {author} {\bibfnamefont {Z.}~\bibnamefont {Guguchia}},
  \bibinfo {author} {\bibfnamefont {R.}~\bibnamefont {Khasanov}}, \bibinfo
  {author} {\bibfnamefont {M.}~\bibnamefont {Chinotti}}, \bibinfo {author}
  {\bibfnamefont {L.}~\bibnamefont {Li}}, \bibinfo {author} {\bibfnamefont
  {K.}~\bibnamefont {Wang}}, \bibinfo {author} {\bibfnamefont {C.}~\bibnamefont
  {Petrovic}},\ and\ \bibinfo {author} {\bibfnamefont {E.}~\bibnamefont
  {Morenzoni}},\ }\bibfield  {journal} {\bibinfo  {journal} {Physical Review
  B}\ }\textbf {\bibinfo {volume} {92}},\ \href
  {https://doi.org/10.1103/physrevb.92.195122} {10.1103/physrevb.92.195122}
  (\bibinfo {year} {2015})\BibitemShut {NoStop}%
\bibitem [{\citenamefont {Ban}\ \emph {et~al.}(2017)\citenamefont {Ban},
  \citenamefont {Luo},\ and\ \citenamefont {Wang}}]{Ban2017}%
  \BibitemOpen
  \bibfield  {author} {\bibinfo {author} {\bibfnamefont {W.~J.}\ \bibnamefont
  {Ban}}, \bibinfo {author} {\bibfnamefont {J.~L.}\ \bibnamefont {Luo}},\ and\
  \bibinfo {author} {\bibfnamefont {N.~L.}\ \bibnamefont {Wang}},\ }\href
  {https://doi.org/10.1088/1361-648x/aa7ef8} {\bibfield  {journal} {\bibinfo
  {journal} {Journal of Physics: Condensed Matter}\ }\textbf {\bibinfo {volume}
  {29}},\ \bibinfo {pages} {405603} (\bibinfo {year} {2017})}\BibitemShut
  {NoStop}%
\bibitem [{\citenamefont {Brydon}\ \emph {et~al.}(2005)\citenamefont {Brydon},
  \citenamefont {Zhu}, \citenamefont {Gul{\'{a}}csi},\ and\ \citenamefont
  {Bishop}}]{Brydon2005}%
  \BibitemOpen
  \bibfield  {author} {\bibinfo {author} {\bibfnamefont {P.~M.~R.}\
  \bibnamefont {Brydon}}, \bibinfo {author} {\bibfnamefont {J.-X.}\
  \bibnamefont {Zhu}}, \bibinfo {author} {\bibfnamefont {M.}~\bibnamefont
  {Gul{\'{a}}csi}},\ and\ \bibinfo {author} {\bibfnamefont {A.~R.}\
  \bibnamefont {Bishop}},\ }\bibfield  {journal} {\bibinfo  {journal} {Physical
  Review B}\ }\textbf {\bibinfo {volume} {72}},\ \href
  {https://doi.org/10.1103/physrevb.72.125122} {10.1103/physrevb.72.125122}
  (\bibinfo {year} {2005})\BibitemShut {NoStop}%
\bibitem [{\citenamefont {Balseiro}\ and\ \citenamefont
  {Falicov}(1979)}]{Balseiro1979}%
  \BibitemOpen
  \bibfield  {author} {\bibinfo {author} {\bibfnamefont {C.~A.}\ \bibnamefont
  {Balseiro}}\ and\ \bibinfo {author} {\bibfnamefont {L.~M.}\ \bibnamefont
  {Falicov}},\ }\href {https://doi.org/10.1103/physrevb.20.4457} {\bibfield
  {journal} {\bibinfo  {journal} {Physical Review B}\ }\textbf {\bibinfo
  {volume} {20}},\ \bibinfo {pages} {4457} (\bibinfo {year}
  {1979})}\BibitemShut {NoStop}%
\bibitem [{\citenamefont {Scalettar}\ \emph {et~al.}(1989)\citenamefont
  {Scalettar}, \citenamefont {Bickers},\ and\ \citenamefont
  {Scalapino}}]{Scalettar1989}%
  \BibitemOpen
  \bibfield  {author} {\bibinfo {author} {\bibfnamefont {R.~T.}\ \bibnamefont
  {Scalettar}}, \bibinfo {author} {\bibfnamefont {N.~E.}\ \bibnamefont
  {Bickers}},\ and\ \bibinfo {author} {\bibfnamefont {D.~J.}\ \bibnamefont
  {Scalapino}},\ }\href {https://doi.org/10.1103/physrevb.40.197} {\bibfield
  {journal} {\bibinfo  {journal} {Physical Review B}\ }\textbf {\bibinfo
  {volume} {40}},\ \bibinfo {pages} {197} (\bibinfo {year} {1989})}\BibitemShut
  {NoStop}%
\bibitem [{\citenamefont {Veki\'{c}}\ \emph {et~al.}(1992)\citenamefont
  {Veki\'{c}}, \citenamefont {Noack},\ and\ \citenamefont {White}}]{Vekic1992}%
  \BibitemOpen
  \bibfield  {author} {\bibinfo {author} {\bibfnamefont {M.}~\bibnamefont
  {Veki\'{c}}}, \bibinfo {author} {\bibfnamefont {R.~M.}\ \bibnamefont
  {Noack}},\ and\ \bibinfo {author} {\bibfnamefont {S.~R.}\ \bibnamefont
  {White}},\ }\href {https://doi.org/10.1103/physrevb.46.271} {\bibfield
  {journal} {\bibinfo  {journal} {Physical Review B}\ }\textbf {\bibinfo
  {volume} {46}},\ \bibinfo {pages} {271} (\bibinfo {year} {1992})}\BibitemShut
  {NoStop}%
\bibitem [{\citenamefont {Das}\ \emph {et~al.}(2008)\citenamefont {Das},
  \citenamefont {Markiewicz},\ and\ \citenamefont {Bansil}}]{Das2008}%
  \BibitemOpen
  \bibfield  {author} {\bibinfo {author} {\bibfnamefont {T.}~\bibnamefont
  {Das}}, \bibinfo {author} {\bibfnamefont {R.~S.}\ \bibnamefont
  {Markiewicz}},\ and\ \bibinfo {author} {\bibfnamefont {A.}~\bibnamefont
  {Bansil}},\ }\bibfield  {journal} {\bibinfo  {journal} {Physical Review B}\
  }\textbf {\bibinfo {volume} {77}},\ \href
  {https://doi.org/10.1103/PhysRevB.77.134516} {10.1103/PhysRevB.77.134516}
  (\bibinfo {year} {2008})\BibitemShut {NoStop}%
\bibitem [{\citenamefont {Onishi}\ and\ \citenamefont
  {Miyake}(2000)}]{Miyake2000}%
  \BibitemOpen
  \bibfield  {author} {\bibinfo {author} {\bibfnamefont {Y.}~\bibnamefont
  {Onishi}}\ and\ \bibinfo {author} {\bibfnamefont {K.}~\bibnamefont
  {Miyake}},\ }\href@noop {} {\bibfield  {journal} {\bibinfo  {journal}
  {Journal of the Physical Society of Japan}\ }\textbf {\bibinfo {volume}
  {69}},\ \bibinfo {pages} {3955} (\bibinfo {year} {2000})}\BibitemShut
  {NoStop}%
\bibitem [{\citenamefont {Alla}\ \emph {et~al.}(2020)\citenamefont {Alla},
  \citenamefont {Alexander}, \citenamefont {Dilipkumar}, \citenamefont
  {Vladimir}, \citenamefont {Erik}, \citenamefont {Yevhen}, \citenamefont
  {Hoon},\ and\ \citenamefont {Sergey}}]{Chikina2020}%
  \BibitemOpen
  \bibfield  {author} {\bibinfo {author} {\bibfnamefont {C.}~\bibnamefont
  {Alla}}, \bibinfo {author} {\bibfnamefont {F.}~\bibnamefont {Alexander}},
  \bibinfo {author} {\bibfnamefont {B.}~\bibnamefont {Dilipkumar}}, \bibinfo
  {author} {\bibfnamefont {V.}~\bibnamefont {Vladimir}}, \bibinfo {author}
  {\bibfnamefont {H.}~\bibnamefont {Erik}}, \bibinfo {author} {\bibfnamefont
  {K.}~\bibnamefont {Yevhen}}, \bibinfo {author} {\bibfnamefont {K.~K.}\
  \bibnamefont {Hoon}},\ and\ \bibinfo {author} {\bibfnamefont
  {B.}~\bibnamefont {Sergey}},\ }\href@noop {} {\bibfield  {journal} {\bibinfo
  {journal} {npj Quantum Materials}\ }\textbf {\bibinfo {volume} {5}},\
  \bibinfo {pages} {1} (\bibinfo {year} {2020})}\BibitemShut {NoStop}%
\bibitem [{\citenamefont {Santos}\ \emph {et~al.}(2010)\citenamefont {Santos},
  \citenamefont {Iglesias}, \citenamefont {Lacroix},\ and\ \citenamefont
  {Gusm{\~{a}}o}}]{santos2010}%
  \BibitemOpen
  \bibfield  {author} {\bibinfo {author} {\bibfnamefont {E.~G.}\ \bibnamefont
  {Santos}}, \bibinfo {author} {\bibfnamefont {J.~R.}\ \bibnamefont
  {Iglesias}}, \bibinfo {author} {\bibfnamefont {C.}~\bibnamefont {Lacroix}},\
  and\ \bibinfo {author} {\bibfnamefont {M.~A.}\ \bibnamefont {Gusm{\~{a}}o}},\
  }\href {https://doi.org/10.1088/0953-8984/22/21/215701} {\bibfield  {journal}
  {\bibinfo  {journal} {Journal of Physics: Condensed Matter}\ }\textbf
  {\bibinfo {volume} {22}},\ \bibinfo {pages} {215701} (\bibinfo {year}
  {2010})}\BibitemShut {NoStop}%
\bibitem [{\citenamefont {Chadi}\ and\ \citenamefont
  {Cohen}(1973)}]{Chadi1973}%
  \BibitemOpen
  \bibfield  {author} {\bibinfo {author} {\bibfnamefont {D.~J.}\ \bibnamefont
  {Chadi}}\ and\ \bibinfo {author} {\bibfnamefont {M.~L.}\ \bibnamefont
  {Cohen}},\ }\href {https://doi.org/10.1103/physrevb.8.5747} {\bibfield
  {journal} {\bibinfo  {journal} {Physical Review B}\ }\textbf {\bibinfo
  {volume} {8}},\ \bibinfo {pages} {5747} (\bibinfo {year} {1973})}\BibitemShut
  {NoStop}%
\bibitem [{\citenamefont {Costa}\ \emph {et~al.}(2018)\citenamefont {Costa},
  \citenamefont {de~Lima}, \citenamefont {Paiva}, \citenamefont {Massalami},\
  and\ \citenamefont {dos Santos}}]{Costa2018}%
  \BibitemOpen
  \bibfield  {author} {\bibinfo {author} {\bibfnamefont {N.~C.}\ \bibnamefont
  {Costa}}, \bibinfo {author} {\bibfnamefont {J.~P.}\ \bibnamefont {de~Lima}},
  \bibinfo {author} {\bibfnamefont {T.}~\bibnamefont {Paiva}}, \bibinfo
  {author} {\bibfnamefont {M.~E.}\ \bibnamefont {Massalami}},\ and\ \bibinfo
  {author} {\bibfnamefont {R.~R.}\ \bibnamefont {dos Santos}},\ }\href
  {https://doi.org/doi.org/10.1088/1361-648X/aaa1ab} {\bibfield  {journal}
  {\bibinfo  {journal} {J. Phys.: Condens. Matter}\ }\textbf {\bibinfo {volume}
  {30}},\ \bibinfo {pages} {045602} (\bibinfo {year} {2018})}\BibitemShut
  {NoStop}%
\bibitem [{\citenamefont {Reyes}\ \emph {et~al.}(2019)\citenamefont {Reyes},
  \citenamefont {Lopes}, \citenamefont {Continentino},\ and\ \citenamefont
  {Thomas}}]{Reyes2019}%
  \BibitemOpen
  \bibfield  {author} {\bibinfo {author} {\bibfnamefont {D.}~\bibnamefont
  {Reyes}}, \bibinfo {author} {\bibfnamefont {N.}~\bibnamefont {Lopes}},
  \bibinfo {author} {\bibfnamefont {M.~A.}\ \bibnamefont {Continentino}},\ and\
  \bibinfo {author} {\bibfnamefont {C.}~\bibnamefont {Thomas}},\ }\href@noop {}
  {\bibfield  {journal} {\bibinfo  {journal} {Physical Review B}\ }\textbf
  {\bibinfo {volume} {99}},\ \bibinfo {pages} {224514} (\bibinfo {year}
  {2019})}\BibitemShut {NoStop}%
\bibitem [{\citenamefont {Kumakura}\ \emph {et~al.}(1996)\citenamefont
  {Kumakura}, \citenamefont {Tan}, \citenamefont {Handa}, \citenamefont
  {Morishita},\ and\ \citenamefont {Fukuyama}}]{Kumakura1996}%
  \BibitemOpen
  \bibfield  {author} {\bibinfo {author} {\bibfnamefont {T.}~\bibnamefont
  {Kumakura}}, \bibinfo {author} {\bibfnamefont {H.}~\bibnamefont {Tan}},
  \bibinfo {author} {\bibfnamefont {T.}~\bibnamefont {Handa}}, \bibinfo
  {author} {\bibfnamefont {M.}~\bibnamefont {Morishita}},\ and\ \bibinfo
  {author} {\bibfnamefont {H.}~\bibnamefont {Fukuyama}},\ }\href
  {https://doi.org/10.1007/BF02570292} {\bibfield  {journal} {\bibinfo
  {journal} {Czechoslovak Journal of Physics}\ }\textbf {\bibinfo {volume}
  {46}},\ \bibinfo {pages} {2611–2612} (\bibinfo {year} {1996})}\BibitemShut
  {NoStop}%
\bibitem [{\citenamefont {Singh}\ \emph {et~al.}(2005)\citenamefont {Singh},
  \citenamefont {Nirmala}, \citenamefont {Ramakrishnan},\ and\ \citenamefont
  {Malik}}]{Singh2005}%
  \BibitemOpen
  \bibfield  {author} {\bibinfo {author} {\bibfnamefont {Y.}~\bibnamefont
  {Singh}}, \bibinfo {author} {\bibfnamefont {R.}~\bibnamefont {Nirmala}},
  \bibinfo {author} {\bibfnamefont {S.}~\bibnamefont {Ramakrishnan}},\ and\
  \bibinfo {author} {\bibfnamefont {S.~K.}\ \bibnamefont {Malik}},\ }\href
  {https://doi.org/10.1103/PhysRevB.72.045106} {\bibfield  {journal} {\bibinfo
  {journal} {Physical Review B}\ }\textbf {\bibinfo {volume} {72}},\ \bibinfo
  {pages} {045106} (\bibinfo {year} {2005})}\BibitemShut {NoStop}%
\bibitem [{\citenamefont {Yutaro}\ \emph {et~al.}(2013)\citenamefont {Yutaro},
  \citenamefont {Nobutaka}, \citenamefont {Akihiro}, \citenamefont {Hideki},
  \citenamefont {Hirofumi}, \citenamefont {Masaki}, \citenamefont {Masahiko},\
  and\ \citenamefont {Yutaka}}]{Nagano2013}%
  \BibitemOpen
  \bibfield  {author} {\bibinfo {author} {\bibfnamefont {N.}~\bibnamefont
  {Yutaro}}, \bibinfo {author} {\bibfnamefont {A.}~\bibnamefont {Nobutaka}},
  \bibinfo {author} {\bibfnamefont {M.}~\bibnamefont {Akihiro}}, \bibinfo
  {author} {\bibfnamefont {Y.}~\bibnamefont {Hideki}}, \bibinfo {author}
  {\bibfnamefont {W.}~\bibnamefont {Hirofumi}}, \bibinfo {author}
  {\bibfnamefont {I.}~\bibnamefont {Masaki}}, \bibinfo {author} {\bibfnamefont
  {I.}~\bibnamefont {Masahiko}},\ and\ \bibinfo {author} {\bibfnamefont
  {U.}~\bibnamefont {Yutaka}},\ }\href@noop {} {\bibfield  {journal} {\bibinfo
  {journal} {Journal of the Physical Society of Japan}\ }\textbf {\bibinfo
  {volume} {82}},\ \bibinfo {pages} {064715} (\bibinfo {year}
  {2013})}\BibitemShut {NoStop}%
\bibitem [{\citenamefont {Huixia}\ \emph {et~al.}(2015)\citenamefont {Huixia},
  \citenamefont {Weiwei}, \citenamefont {Jing}, \citenamefont {Hiroyuki},
  \citenamefont {Andr{\'a}s}, \citenamefont {W}, \citenamefont {Ali},
  \citenamefont {Yimei},\ and\ \citenamefont {Joseph}}]{Luo2015}%
  \BibitemOpen
  \bibfield  {author} {\bibinfo {author} {\bibfnamefont {L.}~\bibnamefont
  {Huixia}}, \bibinfo {author} {\bibfnamefont {X.}~\bibnamefont {Weiwei}},
  \bibinfo {author} {\bibfnamefont {T.}~\bibnamefont {Jing}}, \bibinfo {author}
  {\bibfnamefont {I.}~\bibnamefont {Hiroyuki}}, \bibinfo {author}
  {\bibfnamefont {G.}~\bibnamefont {Andr{\'a}s}}, \bibinfo {author}
  {\bibfnamefont {K.~J.}\ \bibnamefont {W}}, \bibinfo {author} {\bibfnamefont
  {Y.}~\bibnamefont {Ali}}, \bibinfo {author} {\bibfnamefont {Z.}~\bibnamefont
  {Yimei}},\ and\ \bibinfo {author} {\bibfnamefont {C.~R.}\ \bibnamefont
  {Joseph}},\ }\href@noop {} {\bibfield  {journal} {\bibinfo  {journal}
  {Proceedings of the National Academy of Sciences}\ }\textbf {\bibinfo
  {volume} {112}},\ \bibinfo {pages} {E1174} (\bibinfo {year}
  {2015})}\BibitemShut {NoStop}%
\bibitem [{\citenamefont {Li}\ \emph {et~al.}(2017)\citenamefont {Li},
  \citenamefont {Deng}, \citenamefont {Wang}, \citenamefont {Liu},
  \citenamefont {Abeykoon}, \citenamefont {Dooryhee}, \citenamefont {Tomic},
  \citenamefont {Huang}, \citenamefont {B.Warren}, \citenamefont {Bozin},
  \citenamefont {Billinge}, \citenamefont {Sun}, \citenamefont {Zhu},
  \citenamefont {Kotliar},\ and\ \citenamefont {Petrovic}}]{Li2017}%
  \BibitemOpen
  \bibfield  {author} {\bibinfo {author} {\bibfnamefont {L.}~\bibnamefont
  {Li}}, \bibinfo {author} {\bibfnamefont {X.}~\bibnamefont {Deng}}, \bibinfo
  {author} {\bibfnamefont {Z.}~\bibnamefont {Wang}}, \bibinfo {author}
  {\bibfnamefont {Y.}~\bibnamefont {Liu}}, \bibinfo {author} {\bibfnamefont
  {M.}~\bibnamefont {Abeykoon}}, \bibinfo {author} {\bibfnamefont
  {E.}~\bibnamefont {Dooryhee}}, \bibinfo {author} {\bibfnamefont
  {A.}~\bibnamefont {Tomic}}, \bibinfo {author} {\bibfnamefont
  {Y.}~\bibnamefont {Huang}}, \bibinfo {author} {\bibfnamefont
  {J.}~\bibnamefont {B.Warren}}, \bibinfo {author} {\bibfnamefont {E.~S.}\
  \bibnamefont {Bozin}}, \bibinfo {author} {\bibfnamefont {S.~J.~L.}\
  \bibnamefont {Billinge}}, \bibinfo {author} {\bibfnamefont {Y.}~\bibnamefont
  {Sun}}, \bibinfo {author} {\bibfnamefont {Y.}~\bibnamefont {Zhu}}, \bibinfo
  {author} {\bibfnamefont {G.}~\bibnamefont {Kotliar}},\ and\ \bibinfo {author}
  {\bibfnamefont {C.}~\bibnamefont {Petrovic}},\ }\href
  {https://doi.org/10.1038/s41535-017-0016-9} {\bibfield  {journal} {\bibinfo
  {journal} {npj Quantum Materials}\ }\textbf {\bibinfo {volume} {2}},\
  \bibinfo {pages} {1} (\bibinfo {year} {2017})}\BibitemShut {NoStop}%
\bibitem [{\citenamefont {Cho}\ \emph {et~al.}(2018)\citenamefont {Cho},
  \citenamefont {M.Ko{\'n}czykowski}, \citenamefont {Teknowijoyo},
  \citenamefont {Tanatar}, \citenamefont {Guss}, \citenamefont {Gartin},
  \citenamefont {Wilde}, \citenamefont {Kreyssig}, \citenamefont
  {R.~J.~McQueeney}, \citenamefont {Mishra}, \citenamefont {Hirschfeld}, ,\
  and\ \citenamefont {Prozorov}}]{Cho2018}%
  \BibitemOpen
  \bibfield  {author} {\bibinfo {author} {\bibfnamefont {K.}~\bibnamefont
  {Cho}}, \bibinfo {author} {\bibnamefont {M.Ko{\'n}czykowski}}, \bibinfo
  {author} {\bibfnamefont {S.}~\bibnamefont {Teknowijoyo}}, \bibinfo {author}
  {\bibfnamefont {M.~A.}\ \bibnamefont {Tanatar}}, \bibinfo {author}
  {\bibfnamefont {J.}~\bibnamefont {Guss}}, \bibinfo {author} {\bibfnamefont
  {P.~B.}\ \bibnamefont {Gartin}}, \bibinfo {author} {\bibfnamefont {J.~M.}\
  \bibnamefont {Wilde}}, \bibinfo {author} {\bibfnamefont {A.}~\bibnamefont
  {Kreyssig}}, \bibinfo {author} {\bibfnamefont {A.~I.~G.}\ \bibnamefont
  {R.~J.~McQueeney}}, \bibinfo {author} {\bibfnamefont {V.}~\bibnamefont
  {Mishra}}, \bibinfo {author} {\bibfnamefont {P.~J.}\ \bibnamefont
  {Hirschfeld}}, ,\ and\ \bibinfo {author} {\bibfnamefont {R.}~\bibnamefont
  {Prozorov}},\ }\href {https://doi.org/10.1038/s41467-018-05153-0} {\bibfield
  {journal} {\bibinfo  {journal} {Nature Communications}\ }\textbf {\bibinfo
  {volume} {9}},\ \bibinfo {pages} {1} (\bibinfo {year} {2018})}\BibitemShut
  {NoStop}%
\bibitem [{\citenamefont {Khomskii}(2010)}]{Khomskii2010}%
  \BibitemOpen
  \bibfield  {author} {\bibinfo {author} {\bibfnamefont {D.~I.}\ \bibnamefont
  {Khomskii}},\ }\href
  {https://www.ebook.de/de/product/11443694/daniel_i_khomskii_basic_aspects_of_the_quantum_theory_of_solids.html}
  {\emph {\bibinfo {title} {Basic aspects of the quantum theory of solids:
  Order and elementary excitations}}}\ (\bibinfo  {publisher} {Cambridge
  University Press},\ \bibinfo {year} {2010})\BibitemShut {NoStop}%
\bibitem [{\citenamefont {Peter}\ and\ \citenamefont {A}(1964)}]{Fulde1964}%
  \BibitemOpen
  \bibfield  {author} {\bibinfo {author} {\bibfnamefont {F.}~\bibnamefont
  {Peter}}\ and\ \bibinfo {author} {\bibfnamefont {F.~R.}\ \bibnamefont {A}},\
  }\href@noop {} {\bibfield  {journal} {\bibinfo  {journal} {Physical Review}\
  }\textbf {\bibinfo {volume} {135}},\ \bibinfo {pages} {A550} (\bibinfo {year}
  {1964})}\BibitemShut {NoStop}%
\bibitem [{\citenamefont {A.I.}\ and\ \citenamefont
  {Ovchinnikov}(1965)}]{Larkin1965}%
  \BibitemOpen
  \bibfield  {author} {\bibinfo {author} {\bibfnamefont {L.}~\bibnamefont
  {A.I.}}\ and\ \bibinfo {author} {\bibfnamefont {Y.}~\bibnamefont
  {Ovchinnikov}},\ }\href@noop {} {\bibfield  {journal} {\bibinfo  {journal}
  {Sov. Phys. JETP.}\ }\textbf {\bibinfo {volume} {20}},\ \bibinfo {pages}
  {762} (\bibinfo {year} {1965})}\BibitemShut {NoStop}%
\bibitem [{\citenamefont {Agterberg}\ \emph {et~al.}(2020)\citenamefont
  {Agterberg}, \citenamefont {Davis}, \citenamefont {Edkins}, \citenamefont
  {Fradkin}, \citenamefont {Harlingen}, \citenamefont {Kivelson}, \citenamefont
  {Lee}, \citenamefont {Radzihovsky}, \citenamefont {Tranquada},\ and\
  \citenamefont {Wang}}]{Agterberg2020}%
  \BibitemOpen
  \bibfield  {author} {\bibinfo {author} {\bibfnamefont {D.~F.}\ \bibnamefont
  {Agterberg}}, \bibinfo {author} {\bibfnamefont {J.~C.~S.}\ \bibnamefont
  {Davis}}, \bibinfo {author} {\bibfnamefont {S.~D.}\ \bibnamefont {Edkins}},
  \bibinfo {author} {\bibfnamefont {E.}~\bibnamefont {Fradkin}}, \bibinfo
  {author} {\bibfnamefont {D.~J.~V.}\ \bibnamefont {Harlingen}}, \bibinfo
  {author} {\bibfnamefont {S.~A.}\ \bibnamefont {Kivelson}}, \bibinfo {author}
  {\bibfnamefont {P.~A.}\ \bibnamefont {Lee}}, \bibinfo {author} {\bibfnamefont
  {L.}~\bibnamefont {Radzihovsky}}, \bibinfo {author} {\bibfnamefont {J.~M.}\
  \bibnamefont {Tranquada}},\ and\ \bibinfo {author} {\bibfnamefont
  {Y.}~\bibnamefont {Wang}},\ }\href
  {https://doi.org/doi.org/10.1146/annurev-conmatphys-031119-050711} {\bibfield
   {journal} {\bibinfo  {journal} {Annual Review of Condensed Matter Physics}\
  }\textbf {\bibinfo {volume} {11}},\ \bibinfo {pages} {231} (\bibinfo {year}
  {2020})}\BibitemShut {NoStop}%
\bibitem [{\citenamefont {Lopes}\ \emph {et~al.}(2020)\citenamefont {Lopes},
  \citenamefont {Barci},\ and\ \citenamefont {Continentino}}]{Nei2020}%
  \BibitemOpen
  \bibfield  {author} {\bibinfo {author} {\bibfnamefont {N.}~\bibnamefont
  {Lopes}}, \bibinfo {author} {\bibfnamefont {D.~G.}\ \bibnamefont {Barci}},\
  and\ \bibinfo {author} {\bibfnamefont {M.~A.}\ \bibnamefont {Continentino}},\
  }\bibfield  {journal} {\bibinfo  {journal} {J. Phys.: Condens. Matter}\
  }\textbf {\bibinfo {volume} {32}},\ \href
  {https://doi.org/10.1088/1361-648X/ab9a7c} {10.1088/1361-648X/ab9a7c}
  (\bibinfo {year} {2020})\BibitemShut {NoStop}%
\bibitem [{\citenamefont {Farka\v{s}ovsk\'{y}}(2008)}]{Farka2008}%
  \BibitemOpen
  \bibfield  {author} {\bibinfo {author} {\bibfnamefont {P.}~\bibnamefont
  {Farka\v{s}ovsk\'{y}}},\ }\bibfield  {journal} {\bibinfo  {journal} {Physical
  Review B}\ }\textbf {\bibinfo {volume} {77}},\ \href
  {https://doi.org/10.1103/PhysRevB.77.155130} {10.1103/PhysRevB.77.155130}
  (\bibinfo {year} {2008})\BibitemShut {NoStop}%
\bibitem [{\citenamefont {Anurag~Banerjee}\ and\ \citenamefont
  {Ghosal}(2018)}]{Banerjee2018}%
  \BibitemOpen
  \bibfield  {author} {\bibinfo {author} {\bibfnamefont {A.~G.}\ \bibnamefont
  {Anurag~Banerjee}}\ and\ \bibinfo {author} {\bibfnamefont {A.}~\bibnamefont
  {Ghosal}},\ }\bibfield  {journal} {\bibinfo  {journal} {Physical Review B}\
  }\textbf {\bibinfo {volume} {98}},\ \href
  {https://doi.org/10.1103/PhysRevB.98.104206} {10.1103/PhysRevB.98.104206}
  (\bibinfo {year} {2018})\BibitemShut {NoStop}%
\end{thebibliography}%

\end{document}